\documentclass[]{aastex631}
\usepackage[utf8]{inputenc}
\usepackage{graphicx}
\usepackage{natbib}
\usepackage{amsmath}
\usepackage{amssymb}
\usepackage{mathtools}
\usepackage{nccmath}
\usepackage{hyperref}

\hypersetup{
    colorlinks=true,
    linkcolor=blue,
    filecolor=magenta,   
    citecolor=blue,
    urlcolor=cyan,
}

\shorttitle{AASTeX v6.3.1 Sample article}
\shortauthors{Shah et al.}
\graphicspath{{./}{figures/}}

\begin{document}

\title{Interior Structure Models Of Venus.}

\author{Oliver Shah}
\affiliation{Center for Theoretical Astrophysics and Cosmology, University of Zurich, Switzerland}
\affiliation{Center for Space and Habitability, University of Bern, Switzerland}

\author{Ravit Helled}
\affiliation{Center for Theoretical Astrophysics and Cosmology, University of Zurich, Switzerland}

\author{Yann Alibert}
\affiliation{Center for Space and Habitability, University of Bern, Switzerland}

\author{Klaus Mezger}
\affiliation{Center for Space and Habitability, University of Bern, Switzerland}
\affiliation{Institut für Geologie, University of Bern, Switzerland}



\begin{abstract}

  Venus' mass and radius are similar to those of Earth. However, dissimilarities in atmospheric properties, geophysical activity and magnetic field generation could hint towards significant differences in the chemical composition and interior evolution of the two planets.
  Although various explanations for the differences between Venus and Earth have been proposed, the currently available data are insufficient to discriminate among the different solutions. Here we investigate the possible range of Venus structure models. We assume that core segregation happened as a single-stage event. The mantle composition is inferred from the core composition using a prescription for metal-silicate partitioning.
We consider three different cases for the composition of Venus defined via the bulk Si and Mg content, and the core's S content.
  Permissible ranges for the core size, mantle and core composition as well as the normalized moment of inertia (MoI) are presented for these compositions.
  A solid inner core could exist for all compositions.
  We estimate that Venus' MoI is 0.317-0.351 and its core size 2930-4350~km for all assumed  compositions. Higher MoI values correspond to more oxidizing conditions during core segregation. A determination of the abundance of FeO in Venus' mantle by future missions could further constrain its composition and internal structure. This can reveal important information on Venus' formation and evolution, and possibly, the reasons for the differences between Venus and our home planet.  

\end{abstract}



\section{Introduction} \label{sec:intro}

Venus is visible in the night sky by the naked eye and belongs to one of the longest studied objects by astronomers. Due to its similarity to Earth in size, mass, density, and orbital distance it is often referred to as Earth's twin. 
However, its surface conditions are strongly dissimilar to those of Earth. Its atmosphere is about one hundred times denser than Earth's atmosphere and consists predominantly of $\rm CO_2$ and CO while being almost completely devoid of $\rm H_2O$. These discrepancies in atmospheric composition hint to significantly different internal and atmospheric evolution histories of the two planets (e.g. \cite{Gaillard2014}).
Furthermore, unlike Earth, Venus  currently lacks a significant magnetic dynamo in the core and is tectonically inactive (e.g. \cite{Davaille2017}, \cite{Dumoulin2017}, \cite{Stevenson1983}). In addition, Venus has no moon and performs retrograde rotation (e.g. \cite{Raymond2013}, \cite{Gillmann2016}). Moreover, its density seemingly falls out of line with the other terrestrial planets. Although Venus is closer to the Sun, its density is $\sim$2\% lower than predicted if the same bulk composition as Earth is assumed (\cite{Lewis1972}, \cite{Ringwood1977}, \cite{Andersen1980}, \cite{Goettel1981}, \cite{Kaula1990}, \cite{Kaula1994a}, \cite{Basilevsky2003}, \cite{Aitta2012}, \cite{Dumoulin2017}). 

Physical segregation and equilibrium condensation in the solar nebula could lead to a decrease of the intrinsic density of planetary bodies with heliocentric distance (\cite{Kovach1965}, \cite{Lewis1972}, \cite{Ringwood1977}). Such a strict trend, however, is not observed in the solar system. Various mechanisms as possible explanation for the density deficit of Venus have been presented. In particular, it was proposed that Venus' density deficit could be attributed to a lower S content relative to Earth (\cite{Lewis1972}). This idea was challenged since the composition of Venus' atmosphere could hint to an interior that is enriched in S (\cite{Ringwood1977}). It was further pointed out by these authors that Venus' S content as estimated by \cite{Lewis1972} is insufficient to explain its density. The density deficit of Venus was attributed to differences in the oxidation sate of the mantle. This conclusion was based on the assumption that Venus' lower mass lead to overall lower accretion temperatures than for the Earth. 
It was concluded that the [FeO]/[FeO + MgO] ratio of Venus' mantle is about a factor of two higher than that of Earth's mantle (\cite{Ringwood1977}). Consequently, Venus would possess a smaller core than Earth.

\cite{Morgan1980} predicted the composition of the planets in the solar system using a more general framework. It was argued that elements with similar cosmochemical properties exhibit similar fractionation behaviour in the solar nebula. These elements can be divided into five groups, dominantly based on their respective condensation temperature. The bulk composition of a planet can then be determined by knowing the abundances of one element in each of these groups. It was found that Venus could contain less Fe and S than the Earth. This would results in a density that is  $\sim$1.5\% lower than the density of the Earth.

 \citet{Goettel1981}  suggested that the thermal state of the interior played a key role in constraining the planetary composition. Venus'  internal temperature  is expected to be higher than Earth's (\cite{Ringwood1977}, \cite{Goettel1981}, \cite{Stevenson1983}). This is in agreement with more recent estimates based on the composition of basaltic surface rocks on Venus obtained from the Venera 13 and 14 missions in 1981-1982 (\cite{Lee2009}). Higher temperatures and lower pressures inside Venus could be sufficient to have prevented core crystallization which might explain the absence of a magnetic dynamo (\cite{Stevenson1983}, \cite{Stevenson1983a}, \cite{Stevenson2003}, \cite{Aitta2012}, \cite{Jacobson2017}). 
\cite{Goettel1981} proposed that in the absence of subduction the internal temperature might be  high enough to explain Venus' density. However, there is experimental and observational evidence that some form of subduction may occur on Venus (e.g. \cite{Davaille2017}). 

None of these models can be excluded from currently available observational constraints. It was pointed out early on that definite resolution of the ambiguity would require seismic data of Venus' interior (\cite{Goettel1981}). This conclusion was based on the notion that non-hydrostatic effects in the slowly rotating Venus render reliable measurements of the normalized moment of inertia factor $C/MR^2$ (hereafter MoI) very difficult (\cite{Kaula1979}). However, recent progresses in radar observations have opened new possibilities.  \citet{Margot2021a} used radar data, obtained by Earth-based observations in 2006-2020, and estimated Venus' MoI to be $0.337\pm0.024$. This is an important step towards understanding the differences between Earth and Venus. However, the current uncertainty of about 7\% on the MoI is still too large to place firm enough constraints on interior models. Future missions to Venus could drastically improve upon this estimate. EnVision, ESA's candidate mission for the cosmic vision program, is expected to constrain the MoI within $\sim$1.4\% uncertainty (\cite{Rosenblatt2021}).

In this study we investigate possible interior models for Venus.  
In particular, we focus on differences in the chemical composition and the thermal state of the mantle and the core. A single-stage core segregation model is employed to self-consistently compute the mantle composition from the core composition using recent metal-silicate partitioning data for O and Si in multi-component fluids. This approach constrains the light element content of the core, the oxidation state of the mantle, the size of the core, and the presence of a solid inner core.

\section{Model}\label{sec:model}

\subsection{The structure equations}
The planets internal structure is inferred by integrating the one-dimensional structure equations for spherical objects in hydrostatic equilibrium:

 \begin{equation}\label{eq:dPdr}
     \frac{dP}{dr} = -\frac{G m \rho}{r^2},
 \end{equation}
 
 \begin{equation}\label{eq:dmdr}
     \frac{dm}{dr} = 4 \pi \rho r^2,
 \end{equation}

\begin{equation}\label{eq:dTdr}
    \frac{dT}{dr} = \frac{dP}{dr} \left(\frac{dT}{dP}\right)_{ad},
\end{equation}

where $P(r)$, $T(r)$, $\rho(r)$, and $m(r)$ are the pressure, temperature, density, and enclosed mass at radial distance $r$ from the center. $G$ is the gravitational constant and $(dT/dP)_{ad}$ the adiabatic gradient. To close the system of differential equations eq.~\ref{eq:dPdr}-\ref{eq:dTdr} an equation of state (EoS) of the form $\rho(r) = \rho(T(r), P(r))$ for all the constituent materials is required. These EoS are described in Appendix \ref{app:theos}. A 4th order Runge-Kutta scheme was then employed from the center outward to obtain the pressure, temperature and density profiles within a planet. The (non-normalized) moment of inertia $C$ is computed according to:

\begin{equation}\label{eq:dCdr}
    \frac{dC}{dr} = \frac{8}{3} \pi \rho(r) r^4.
\end{equation}

\subsection{Bulk composition and boundary conditions}\label{sec:bulk_comp}

\begin{figure}[t]
\centering
\includegraphics[width=\textwidth]{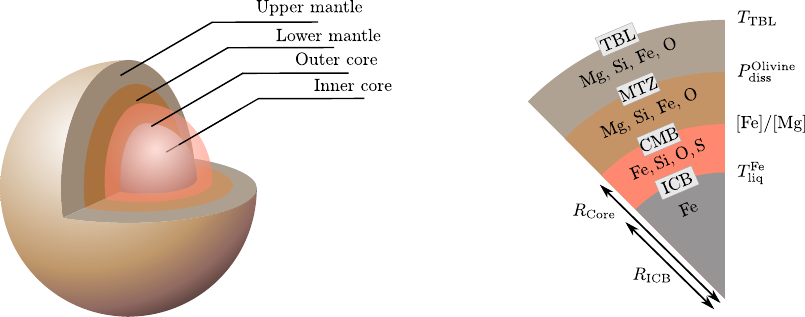}\caption{Sketches of Venus' interior structure. Left: The planetary models are divided into four distinct layers: Upper mantle, lower mantle, liquid outer core, and solid inner core. Right: Composition of each layer and layer transitions. The parameters used to define each layer transition are indicated in the right and are further explained in the main text.}
\label{fig:venus_stat}
\end{figure}

The planetary composition is similar to that presented in \cite{Shah2021}. The interior is divided into four distinct layers (see Fig.~\ref{fig:venus_stat}): An inner solid and an outer liquid iron core, a lower mantle and an upper mantle. The lower mantle consists of the minerals $\rm (Mg, Fe)SiO_3$ (perovskite) and (Mg,Fe)O (wüstite) while the minerals considered in the upper mantle are $\rm (Mg,Fe)_2 SiO_4$ (olivine) and $\rm (Mg,Fe)_2 Si_2 O_6$ (pyroxene) with variable Fe content. This allows a variable [Si]/[Mg] ratio in the mantle. In the outer core S, O and Si are considered as possible impurities. The Earths core is thought to contain other elements such as Ni, C, and H (e.g. \cite{Hirose2021}). These elements were not considered here. This is justified as it was shown that accounting for Si and O is sufficient to reproduce the Earth's mantle composition from the core segregation model (see section~\ref{sec:core_seg_icb}). Furthermore, if applied to the Earth our model is capable of reproducing the content of impurities in the core other than Si, S and O reported in \cite{Hirose2021} if some of the S is replaced by these elements based on mass balance (see Appendix \ref{app:earth_reference} for details). The transition between the upper and lower mantle in the mantle transition zone (MTZ) is determined by the phase transition from olivine to perovskite and wüstite taken at $P_{\rm diss}^{\rm olivine}=$ 25 GPa. The value for the Grüneisenparameter in the core is set to $\gamma = 1.36$ (\cite{Sotin2007}). Details on the EoS of the different materials considered and the numerical procedures employed to solve for the interior structure are given in Appendix~\ref{app:theos} and \cite{Shah2021}. The model accounts for the separation of the core into an inner solid part and an outer liquid part using the melting curve of the Fe-Si-O-S chemical system (see Section \ref{sec:core_seg_icb}). The core mantle boundary (CMB) is defined via the core mass. 
The mole fractions of $\rm SiO_2$ and FeO in the mantle were obtained from chemical equilibrium between mantle and core using experimental data for metal-silicate partitioning of O and Si (see Section \ref{sec:mantle_core_comp}).

The boundary conditions are defined via the surface pressure $P_{\rm S}$, the temperature at the thermal boundary layer (TBL) at the top of the mantle $T_{\rm TBL}$, the Mg-number $\rm Mg \# \equiv$ [Mg]/[Fe + Mg] and the total planetary mass $M$. Eq.~\ref{eq:dPdr}-\ref{eq:dTdr} for a given set of boundary conditions are solved is the same way as in \cite{Shah2021}. The main degrees of freedom in the present model are: the core segregation pressure $P_{\rm CS}$ (see section \ref{sec:mantle_core_comp}), the temperature at the thermal boundary layer $T_{\rm TBL}$, the temperature discontinuity across the mantle transition zone $\Delta T_{\rm MTZ}$, the temperature discontinuity across the core-mantle-boundary (CMB) $\Delta T_{\rm CMB}$, and the amount of FeO, FeS and FeSi in the core, quantified by the mole fractions $X_{\rm FeO}^{\rm Core}$, $X_{\rm FeS}^{\rm Core}$, $X_{\rm FeSi}^{\rm Core}$.

The corresponding value ranges that were considered in this  study are summarized in Table \ref{tab:input_parameter_ranges} (see also Table~\ref{tab:parameter_overview} for an overview of all relevant parameters). For Earth, the  core segregation pressure was inferred to be  30-70 GPa (\cite{Fischer2015}, \cite{Siebert2011}, \cite{Wade2005},\cite{Wood2006}, \cite{Li1996}). Given the similar masses of Earth and Venus, and in order to allow for a direct comparison between the two planets, we use the same range for Venus. The sensitivity of the results on the chosen range of the core segregation pressure is presented in  Appendix~\ref{app:effect_P_CS}. The pressure and temperature conditions of basaltic magma generation on terrestrial planets can be constrained from the composition of surface basalts which allows inferring the temperature at the TBL (\cite{Lee2009}). These authors have used basalts found on Earth and data obtained from the Venera landers to estimate a range of  $\approx1400-1800$ K for the Earth and $\approx1600-1800$ K for Venus. Due to the sparse availability of compositional data for Venus the range for the TBL temperature of Venus could be larger. As a result,  
we consider a range of $T_{\rm TBL}$ between  1400 and 1800 $ K$ . The temperature change across the mantle transition zone of Earth, $\Delta T_{\rm MTZ}$, was estimated to be  $\sim$ 300 K (e.g. \cite{Sotin2007} and references therein). This result corresponds to the assumption that the MTZ arises due to a transition between two different convective regimes in the mantle. However, strong evidence points towards whole-mantle convection within the Earth (e.g. \cite{Morgan1993}, \cite{Arnould2018}, \cite{Loper1985}). Therefore we do not consider a temperature discontinuity across the MTZ  in this study. The temperature drop across the CMB, $\Delta T_{\rm  CMB}$, for Earth was estimated to be 500--1800 K (\cite{Lay2008}). Another study proposed a simple scaling law to compute the temperature drop as $\Delta T_{\rm CMB} \sim 1400 \left(M/M_\oplus\right)$ (\cite{Stixrude2014}). In order to introduce the maximum variability into our model, we consider the wide range of 500--1800 K. No models with $X_{\rm FeO}^{\rm Core}>0.2$ that matched the boundary conditions for Venus were found. Therefore, for the results presented below, a range for $X_{\rm FeO}^{\rm Core}$ between 0 and 0.2 was probed. The Si content of the metal in the silicate-metal partitioning experiments of \cite{Fischer2015} was $\leq$9 wt\%. This corresponds to a mole fraction of FeSi in the core of $\sim$0.25. Here we have set a larger range of 0-0.3 for the maximum FeSi content in the core. The range for the S content in the core is $X_{\rm FeS}^{\rm Core}=0-0.5$ and was constrained from the solar ratio [S]/[Fe] assuming most of the S and Fe to be present in the core (\cite{Lodders2019}).

Due to the very low abundance or even absence of water on Venus it is expected to have only developed a basaltic crust (density $\sim$2.9-3.1 $\rm g \ cm^{-3}$) while the Earth with its higher water content developed also a granitic continental crust (density of $\sim$ 2.7-2.9 $\rm g \ cm^{-3}$) (\cite{Rudnick2018}, \cite{Stolper1980}, \cite{Smrekar2018a}, \cite{Moore2001}). Therefore, the density contrast between the mantle and the crust is likely smaller on Venus than on Earth. As a result, the crust is not explicitly modelled in this study but implicitly incorporated into the upper mantle. Whether granitic regions in the crust are present on Venus (indicating the presence of water in the past) is an open question and could be resolved by future space missions.

The initial values for these parameters are chosen by uniform sampling within the selected ranges for each interior model. Depending on the specific combination of $P_{\rm CS}$, $X_{\rm FeO}^{\rm Core}$, $X_{\rm FeS}^{\rm Core}$, and $X_{\rm FeSi}^{\rm Core}$ it is possible to obtain unrealistic values for the FeO and $\rm SiO_2$ content in the mantle. Here the FeO content in the mantle is limited such that $\rm[FeO]/[FeO + MgO] < 0.5$. The Si-number of the mantle is defined as $\rm Si\#_{Mantle} \equiv [SiO_2]/[MgO+SiO_2]$. It is computed from the $\rm SiO_2$ content according to ${\rm Si \#_{Mantle}} = X_{\rm SiO_2}^{\rm Mantle}/(1-X_{\rm FeO}^{\rm Mantle})$. For the silicate composition in the upper and lower mantle considered in this study ${\rm Si\#}_{\rm Manlte}$ must be between $1/(3-2{\rm Fe}\#_{\rm Mantle})$ and $1/(2-{\rm Fe}\#_{\rm Mantle})$ where $\rm Fe\#_{Mantle} \equiv [FeO]/[FeO+MgO]$. In each numerical run the sampling of the initial parameters is re-iterated until these conditions are met.

 The surface pressure is fixed to 1 bar for Earth and 100 bar for Venus. Variations in the surface pressure of a few tens of bars have no impact on the results presented in this study. The bulk composition was defined as $0.47 \leq \rm Mg\# \leq 0.53$ and $0.47 \leq \rm Si\# \leq 0.56$ (\cite{Sotin2007}). Three different cases for the core composition defined via the S content were considered. A nominal composition was defined as $X^{\rm Core}_{\rm FeS}$ = 0.08-0.15 (see Appendix \ref{app:earth_reference} for details). In order to assess the effect of different S contents on the possible models for Venus S-rich models are defined as $X^{\rm Core}_{\rm FeS}$ = 0.2-0.5 and S-free models as $X^{\rm Core}_{\rm FeS}$ = 0. Convergence of the structure integration is secured via the total mass $M$, $\rm Mg\#$, and $T_{\rm TBL}$: only models for which these parameters match within 0.1\%, 0.1\% and 1\% are considered.

\subsection{Single-stage core segregation and inner core boundary}\label{sec:core_seg_icb}

We adopt the simplifying assumption that chemical equilibrium between the core and the mantle during core segregation is a single-stage process (e.g. \cite{Fischer2015}, \cite{Suer2017}, \cite{Rubie2011}, \cite{Schaefer2017}). This allows omitting explicit modelling of the temporal evolution of the cooling magma ocean. Such a single-stage core formation model was found to yield the correct concentrations of $\rm SiO_2$ and FeO of the Earth's mantle (\cite{Rubie2011}, \cite{Fischer2015}). However, in these studies some trace elements in the mantle such as Ni, Co, W, and Cr could not be reproduced. 
Our study focuses on the abundances of the major elements Si, Mg, O and Fe using the partitioning data from \cite{Fischer2015} and does not consider the trace elements mentioned above in the mantle. We apply the model to Earth as a reference and confirm that the $\rm SiO_2$ and FeO content of the mantle can be reproduced with this model (see Appendix \ref{app:earth_reference}).

\begin{figure}[t]
\centering
\makebox[0pt]{\includegraphics[width=\textwidth]{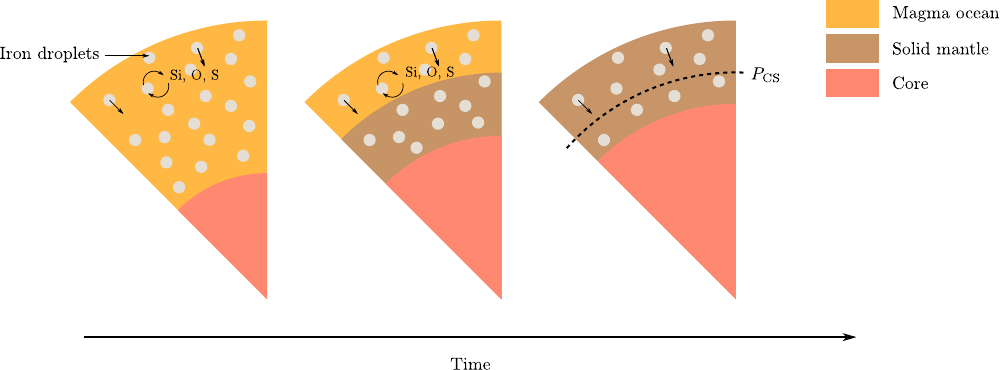}}\caption{A schematic illustration of the core segregation process. The iron core is formed from sinking iron droplets with dissolved Si, O, and S. The Si, O, and S content of the droplets is dictated by the solubility of these elements in liquid iron at the corresponding temperature and pressure conditions. As the mantle solidifies, the lowermost pressure level at which equilibration takes place moves outward. This pressure corresponds to the liquidus of the silicates. In the single stage core segregation model, the equilibration is assumed to have taken place at an average pressure of $P_{\rm CS}$ which is treated as a free parameter.}
\label{fig:core_segregation_illustration}
\end{figure}

The temperature at which equilibrium is reached in the model is given by the pyrolite-liquidus. It is described in its general form using the Simon-Glatzel equation for the melting temperature of solids (\cite{Simon1929}):
\begin{equation}\label{eq:pyrolite_liquidus}
T_{\rm melt}(P)= T_0  \left(\frac{P-P_0}{a}+1\right)^{1/b}. 
\end{equation}
\begin{table}[t]
  \begin{center}
    \caption{Parameters to compute the melting temperature of pyrolite and pure iron from eq.~\ref{eq:pyrolite_liquidus}.}
    \begin{tabular}{lcccc}
      \hline
       Material & $a$ & $b$ & $T_0$ & $P_0$ \\
       & [GPa] & & [K] & [bar] \\
      \hline
         Iron$^{(a)}$ & 23 & 2.26 & 1811 & 1 \\
         Pyrolite$^{(b)}$ & 29 &  1.9 & 1940 & 0 \\

    \hline
    \textbf{References.} $^{(a)}$\cite{Li2020}, $^{(b)}$\cite{Andrault2011}.
    \end{tabular}
    \label{tab:simon_glatzel_parameter}
  \end{center}
\end{table}
In the single-stage core segregation model it is assumed that the chemical equilibration of sinking iron droplets building up the core occurred at an average pressure of $P_{\rm CS}$ (see Fig.~\ref{fig:core_segregation_illustration} for illustration). For a given value of $P_{\rm CS}$ the core segregation temperature $T_{\rm CS}$ is computed via eq.~\ref{eq:pyrolite_liquidus} using the parameters from Table \ref{tab:simon_glatzel_parameter}. The inner core boundary (ICB) is given by the melting curve of the Fe-Si-S-O system described by eq.~\ref{eq:pyrolite_liquidus} and the parameters for pure-iron given in Table \ref{tab:simon_glatzel_parameter}. The effect of impurities on the melting curve in the system are taken from \cite{Andrault2016}. In particular, the melting temperature of the system is reduced by 30 K, 50 K, and 100 K for every weight percent of Si, O, and S, respectively, dissolved in the metal melt.

\begin{figure}[t]
\centering
\makebox[0pt]{\includegraphics[width=\textwidth]{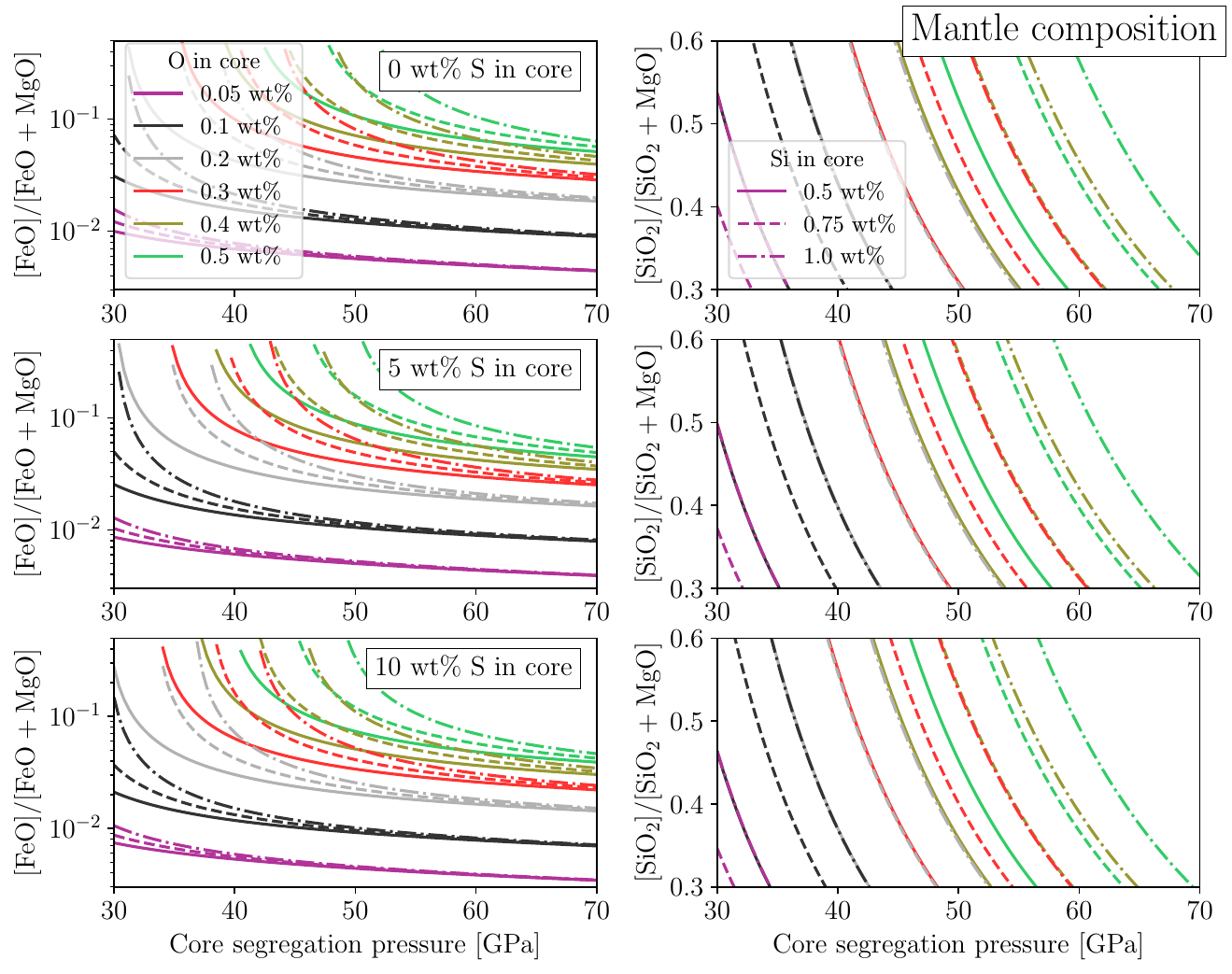}}\caption{FeO and $\rm SiO_2$ content in the mantle as function of the core segregation pressure $P_{\rm CS}$ computed from the metal-silicate partitioning model given by eqs.~\ref{eq:xi_SiO2} and \ref{eq:xi_FeO} for different core compositions. The colors denote different O contents, the line styles different Si contents, and the three rows correspond to 0, 5 and 10 wt\% of S in the core, respectively.}
\label{fig:mantle_composition}
\end{figure}

\begin{figure}[t]
\centering     
\makebox[0pt]{\includegraphics[width=\textwidth]{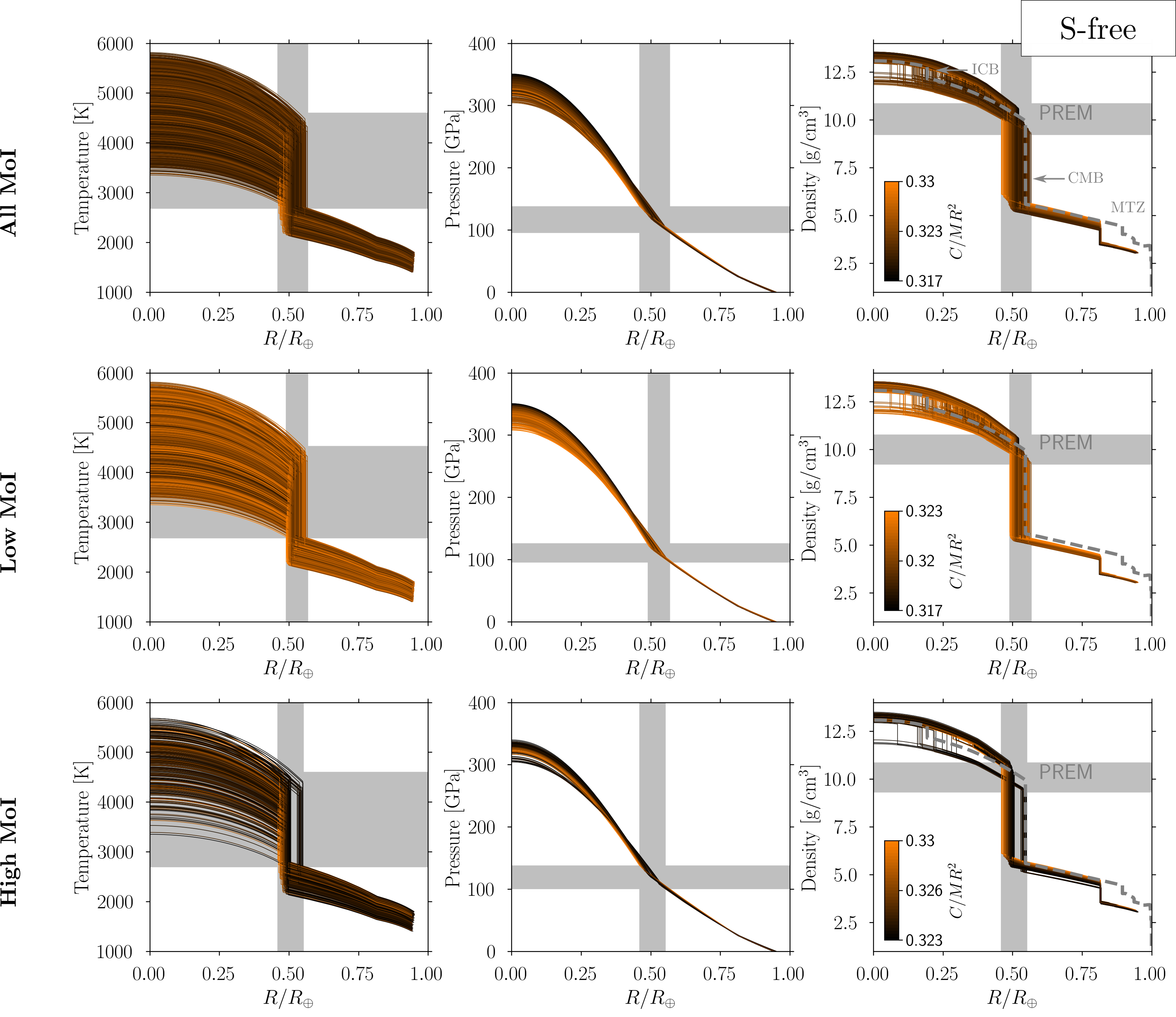}}\caption{Internal temperature (left), pressure (center), and density (right) profiles of the S-free Venus models ($X_{\rm FeS}^{\rm Core} = 0$) as a function of the radial distance from the center $R$ in units of earth radii $R_\oplus$. The top row shows all possible profiles for the allowed range of the MoI. The middle and bottom row show only the profiles within $\pm 1\%$ of the lowest and highest possible value of MoI. The MoI value is indicated by different colors. The grey shaded regions indicate the possible ranges for the size of the core along the x-axes and  the corresponding temperature, pressure and density ranges along the y-axes at the CMB. The dashed lines denote the density profile for the Earth from the Preliminary Earth Reference Model (PREM).}
\label{fig:venus_profiles_sulfur_poor}
\end{figure}

\begin{figure}[t]
\centering
\makebox[0pt]{\includegraphics[width=\textwidth]{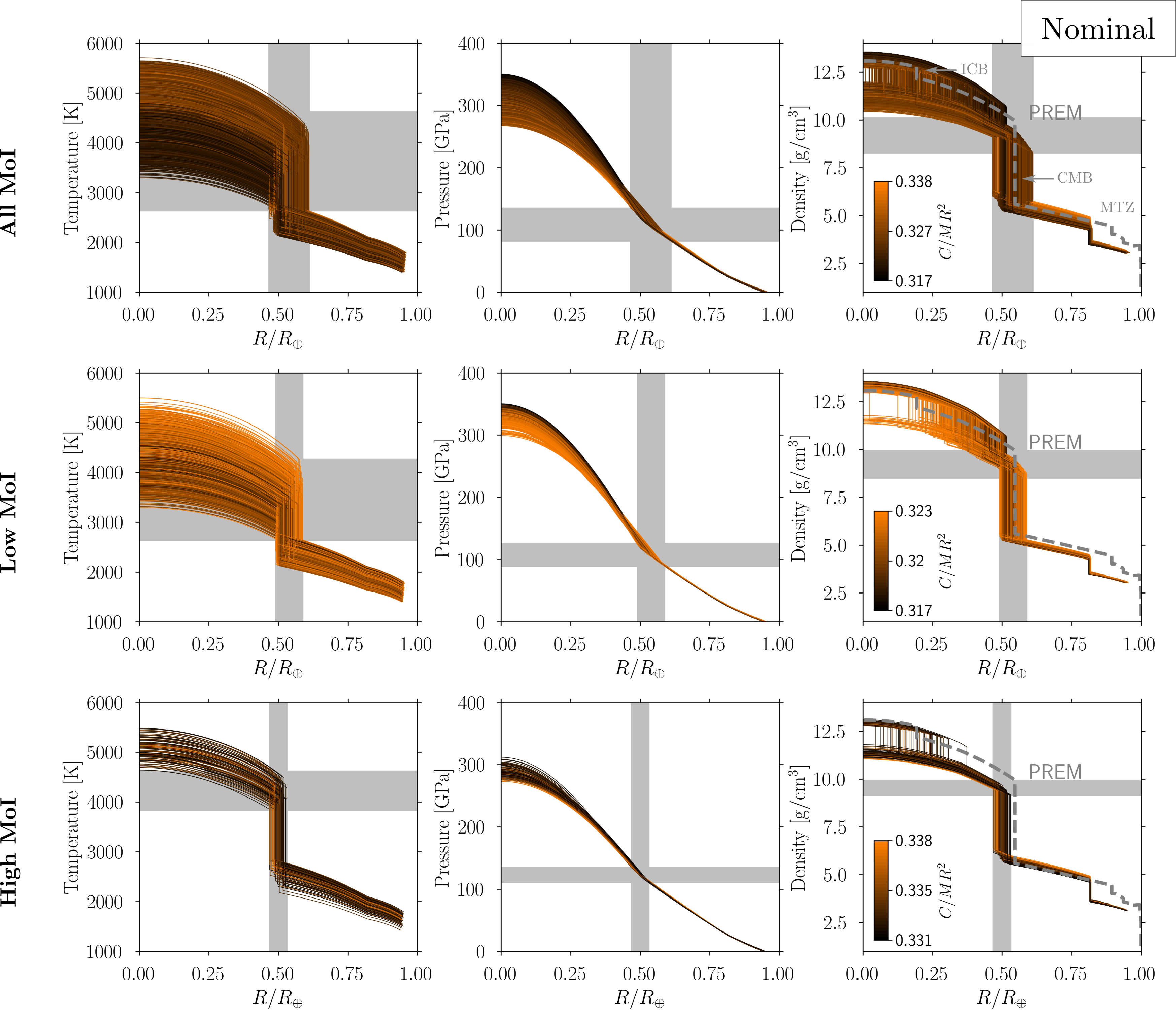}}\caption{Same as Fig.~\ref{fig:venus_profiles_sulfur_poor} but for the nominal Venus models ($X_{\rm FeS}^{\rm Core} = 0.08-0.15$).}
\label{fig:venus_profiles_earth_like}
\end{figure}

\begin{figure}[t]
\centering
\makebox[0pt]{\includegraphics[width=\textwidth]{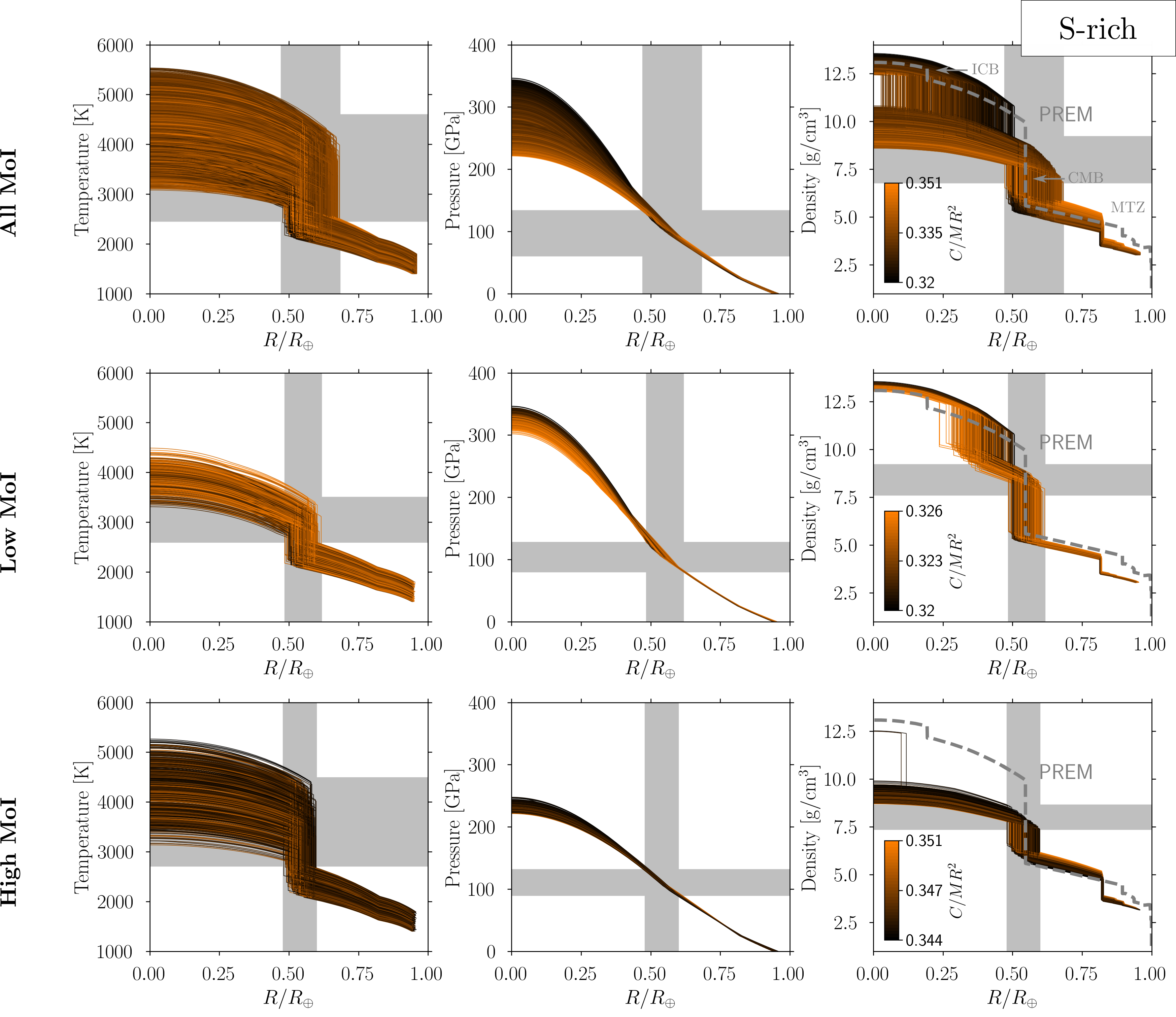}}\caption{Same as Fig.~\ref{fig:venus_profiles_earth_like} but for the S-rich Venus models ($X_{\rm FeS}^{\rm Core} = 0.2-0.5$).}
\label{fig:venus_profiles_sulfur_rich}
\end{figure}

\begin{figure}[t]
\centering
\makebox[0pt]{\includegraphics[width=\textwidth]{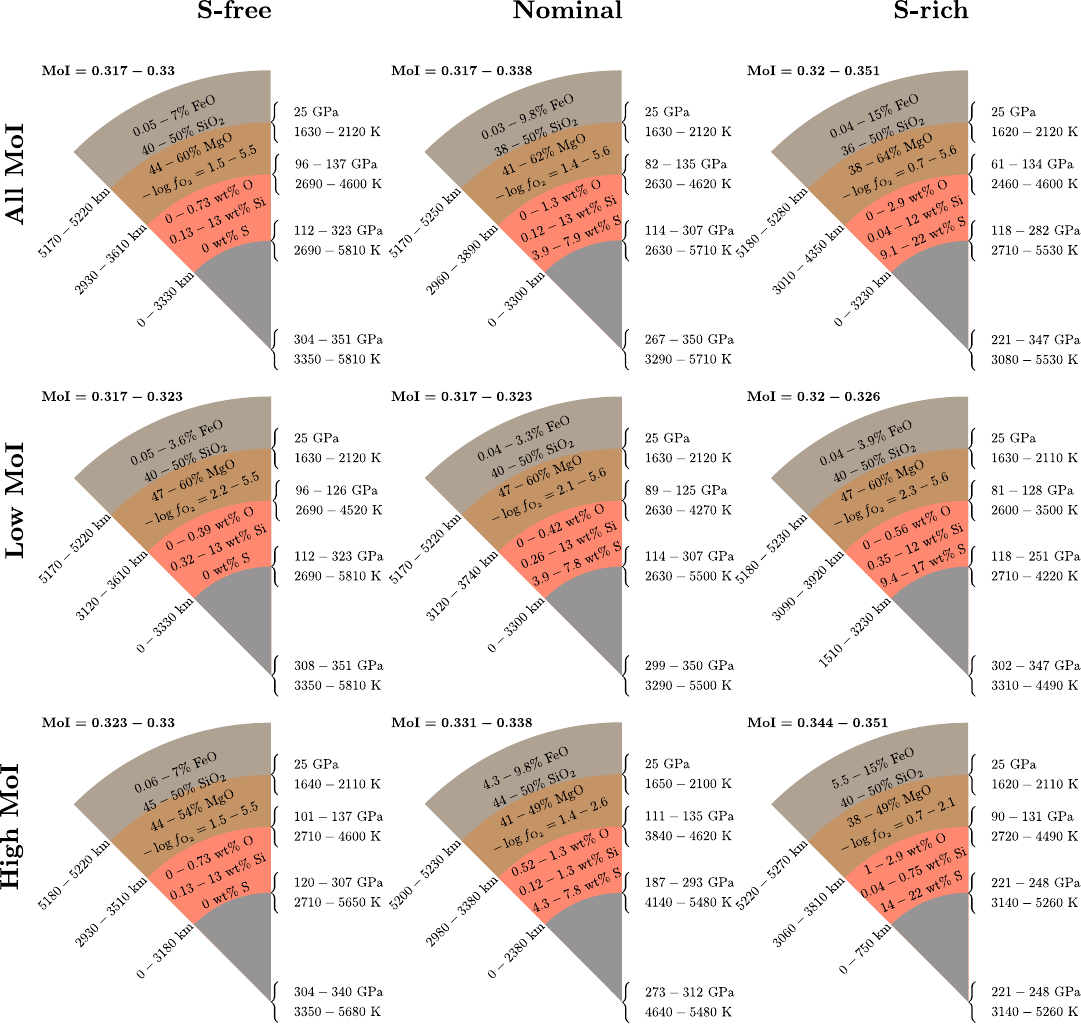}}\caption{Overview of the permissible value ranges of the most relevant structure and composition parameters for the S-free (left), nominal (center), and S-rich (right) models. The pressures and temperatures given for each layer transition correspond to the values in the lower of the two layers at the boundary. The top row shows all possibilities within the permissible MoI range. The middle and bottom row only show the models within $\pm 1\%$ of the lowest and highest possible MoI value. The pressure and temperature values at each layer transition correspond to the values in the layer below the boundary.}
\label{fig:venus_all_scenarios}
\end{figure}

\begin{figure}[t]
\centering
\makebox[0pt]{\includegraphics[width=\textwidth]{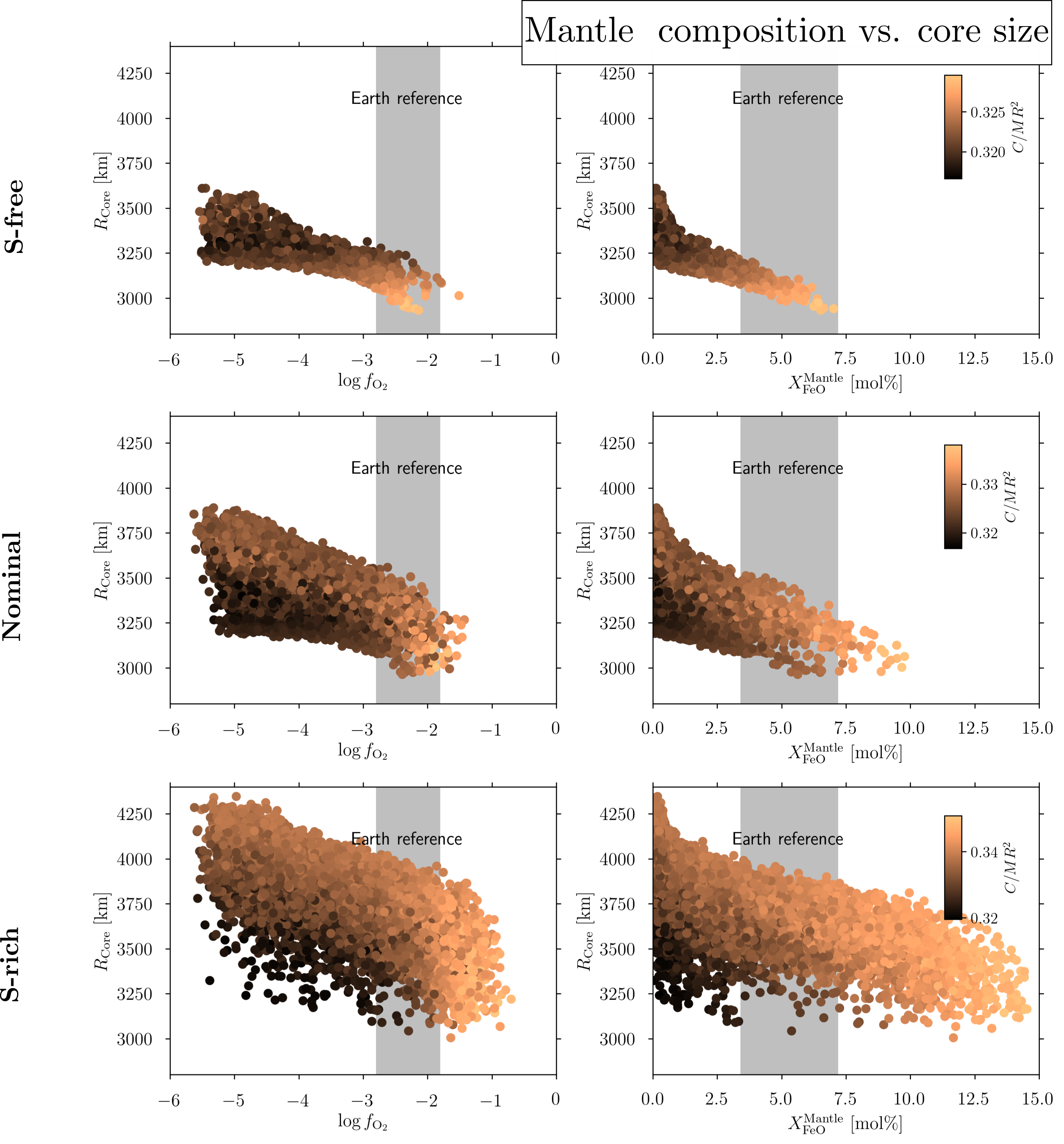}}\caption{Core radius as a function of the redox state of the mantle (left) and the FeO content in the mantle (right) for the different compositions. The value for the MoI is indicat ed by different colors. The permissible ranges of $\log{f_{\rm O_2}}$ relative to the IW buffer and $X_{\rm FeO}^{\rm Mantle}$ for the Earth estimated from our model are shown as grey shaded regions. Even without further constraining the MoI, the core size of Venus could be constrained from the mantle composition alone.}
\label{fig:venus_mantle_comp_vs_core_size}
\end{figure}

\begin{figure}[t]
\centering
\makebox[0pt]{\includegraphics[width=\textwidth]{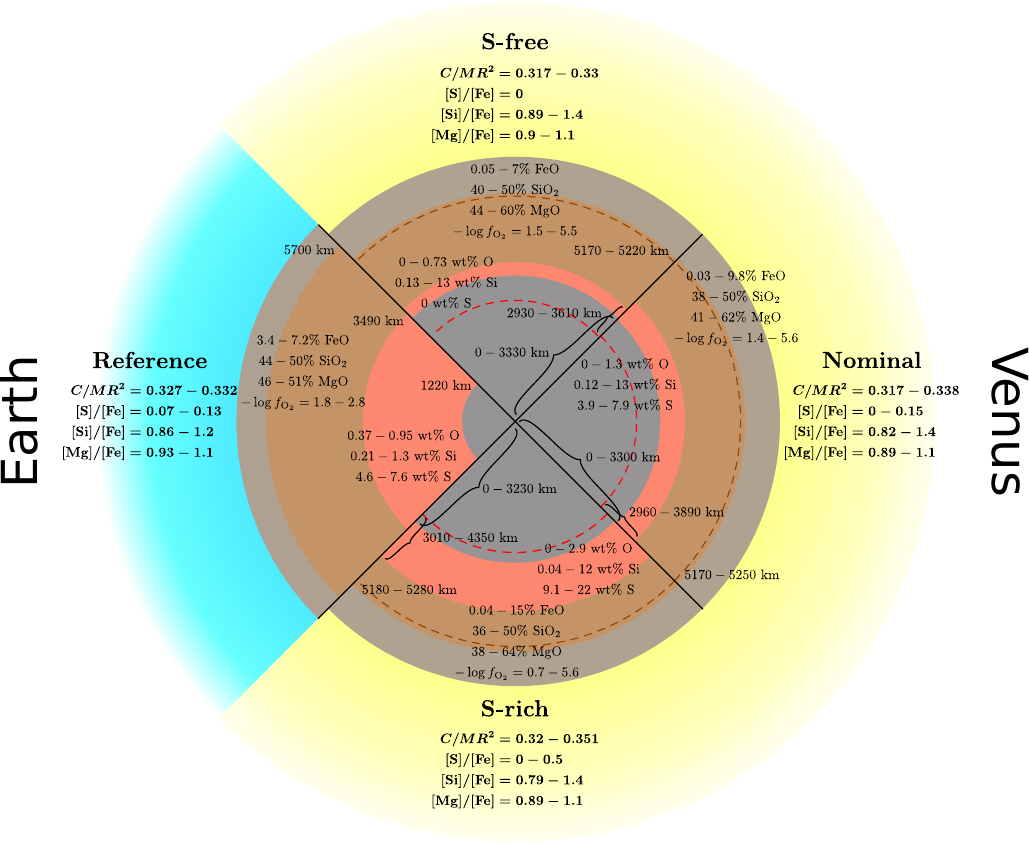}}\caption{A schematic comparison between Earth and the different interior models of Venus. The S-free models correspond to $X_{\rm FeS}^{\rm Core} = 0$ while the nominal and S-rich compositions are defined as $X_{\rm FeS}^{\rm Core} = 0.08-0.15$ and $X_{\rm FeS}^{\rm Core} = 0.2-0.5$. The distances from the center for each layer and the core and mantle compositions are indicated for each case. The brown, red, and black dashed segments indicate the lower bound for the mantle transition zone, the core-mantle boundary and the inner-to-outer core boundary. The relative sizes of the different layers are on scale. The resulting permissible ranges for the MoI and some bulk elemental ratios are given next to the each sketch of the interior.}
\label{fig:Earth_vs_Venus}
\end{figure}

\begin{figure}[t]
\centering
\includegraphics[width=\textwidth]{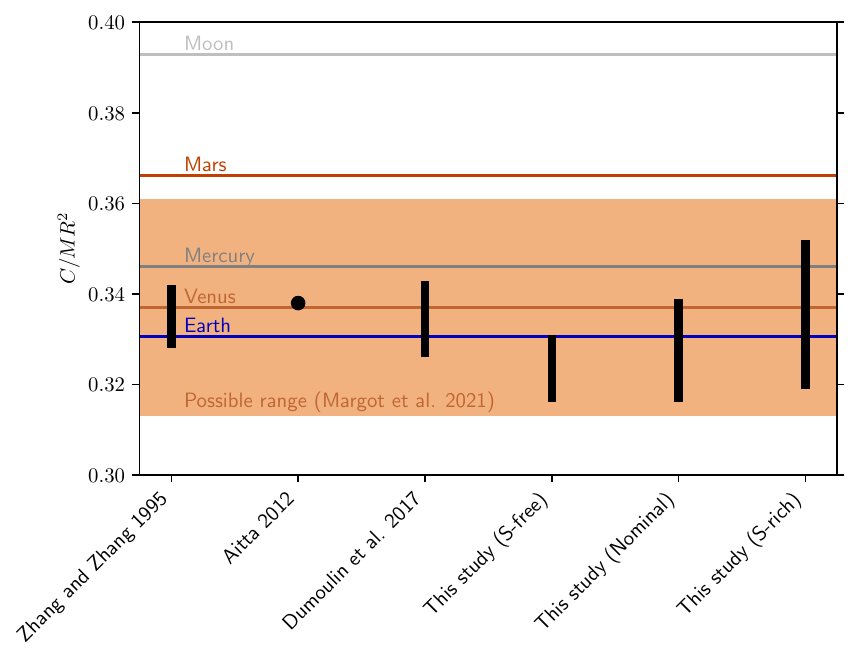}\caption{Estimated ranges of Venus' MoI from \cite{Zhang1995}, \cite{Aitta2012}, and \cite{Dumoulin2017} and the three different bulk compositions we consider. Measured values for the MoI for Earth, Venus, Mercury, and Mars are shown for comparison. The shaded region denotes the current uncertainty of Venus' MoI from observations (\cite{Margot2021a}).}
\label{fig:MoI_ranges}
\end{figure}

\subsection{Mantle and core composition}\label{sec:mantle_core_comp}

\begin{table}
  \begin{center}
    \caption{Parameters for eq.~\ref{eq:KD_mix} and \ref{eq:epsilon} from \cite{Fischer2015}.}
    \begin{tabular}{lccccc}
      \hline
       Element $i$ & $e_{\rm O}^i$ & $e_{\rm Si}^i$ & $a_i$ & $b_i$ [K] & $c_i$ [K/GPa] \\
      \hline
         Si & -0.06 & ... & 0.6 & -11,700 & ... \\
         O & -0.12 & -0.11  & 0.1 & -2200 & 5 \\

    \hline
    \end{tabular}
    \label{tab:epsilon_modelling_coefficients}
  \end{center}
\end{table}

The results for metal-silicate partitioning of O and Si from \cite{Fischer2015} are used to compute the mantle composition from the core composition. Chemical equilibrium between the core and the mantle is assumed to have been achieved in a single-stage process. The mantle composition was defined via the mole fractions $X_{\rm FeO}^{\rm Mantle}$ and $X_{\rm SiO_2}^{\rm Mantle}$:

\begin{equation}\label{eq:xi_SiO2}
X_{\rm SiO_2}^{\rm Mantle} = \frac{X_{\rm Si}^{\rm Core}}{K_{\rm D}^{\rm Si}} \frac{X_{\rm FeO}^{\rm Mantle}}{X_{\rm Fe}^{\rm Core}}
\end{equation}

\begin{equation}\label{eq:xi_FeO}
X_{\rm FeO}^{\rm Mantle} = \frac{X_{\rm Fe}^{\rm Core}X_{\rm O}^{\rm Core}}{K_{\rm D}^{\rm O}},
\end{equation}

where $K_{\rm D}^{\rm Si}$ and $K_{\rm D}^{\rm O}$ are the partition coefficients of O and Si. In a mixture of $N$ components the partition coefficient of component $i$ is given by (\cite{Fischer2015}, \cite{Ma2001}):

\begin{equation}\label{eq:KD_mix}
\begin{split}
\log K_{\rm D}^i(T,P) & = a_i + \frac{b_i}{T} + \frac{c_i P}{T} + \frac{\epsilon_i^i \ln{\left(1-X_i\right)}}{2.303} \\
& + \frac{1}{2.303} \sum_{k=2, k\neq i}^{N} \epsilon_k^i X_k\left(1 + \frac{\ln{\left(1-X_k\right)}}{X_k}- \frac{1}{1-X_i}\right) \\
& - \frac{1}{2.303} \sum_{k=2, k\neq i}^{N} \epsilon_k^i X_k^2 X_i \left(\frac{1}{1-X_i}+\frac{1}{1-X_k} + \frac{X_i}{2\left(1-X_i\right)^2} -1\right).
\end{split}
\end{equation}
The parameter $\epsilon_k^i$, is given by:

\begin{equation}\label{eq:epsilon}
\epsilon_k^i(T) = 1+\frac{e_k^i m_i }{0.242}\frac{T_{\rm ref}}{T}-\frac{m_i}{55.85}
\end{equation}

\begin{table}[ht]
  \begin{center}
    \caption{Probed ranges for the input parameters for the Venus models.}
    \begin{tabular}{lccccccc}
      \hline
       Mg\#  & $P_{\rm CS}$ & $T_{\rm TBL}$ & $\Delta T_{\rm CMB}$ & $X_{\rm FeO}^{\rm Core}$ & $X_{\rm FeS}^{\rm Core}$ & $X_{\rm FeSi}^{\rm Core}$ \\
        & [GPa] & [K] & [K] & & & \\
      \hline
         {0.47-0.53} & 30--70 & {1400-1800} & {500--1800} & 0-0.2 & {0-0.5} & 0-0.3\\
    \hline
    \end{tabular}
    \label{tab:input_parameter_ranges}
  \end{center}
\end{table}

where $m_i$ is the molar mass of species $i$ in grams per mole, $T_{\rm ref} = 1873$ K, and $e_k^i$ is an interaction parameter between species $i$ and $k$. The corresponding parameters are given in Table~\ref{tab:epsilon_modelling_coefficients}. With this the mole fractions of FeO and $\rm SiO_2$ in the mantle are uniquely constrained from the core composition and the conditions at the bottom of the magma ocean during core segregation. The presence of S in the core could impact the partitioning behaviour of Si and O (\cite{Fischer2015}). However, to our knowledge, no self-consistent prescription for these effects are currently available. In order to study the influence of S in the Fe-Si-O-S system a series of challenging experiments would be required (pers. comm. D. Frost). The importance of such experiments was already noted by \cite{Fischer2015}. Incorporating the effect of S into the partitioning model could reduce the redundancy of possible interior models for Venus and should be investigated in the future.

Fig. \ref{fig:mantle_composition} shows the mantle composition resulting from eq.~\ref{eq:xi_SiO2} and \ref{eq:xi_FeO}. The different colors correspond to different O contents in the core while the different linestyles denote different values for the Si content. The three rows correspond to 0, 5 and 10 wt\% S. The oxygen fugacity of the manlte is estimated as a function of temperature and pressure using a mineral redox buffer. Here the iron-wüstite buffer (IW) was used. It is based on the equlibrium reaction $2 {\rm Fe} + {\rm O}_2 = 2 {\rm Fe} {\rm O}$. The oxygen fugacity is typically given in log units relative to a chosen redox buffer. Negative values correspond to oxygen fugacities that are lower than the equilibrium of the buffer at a given temperature and pressure. The oxygen fugacity relative to the iron-wüstite buffer is calculated as (\cite{Weston2009}):

\begin{equation}\label{eq:logfO2}
\log f_{\rm O_2} = 2 \log \left(\frac{a_{\rm FeO}^{\rm sil}}{a_{\rm Fe}^{\rm met}}\right),
\end{equation}

where:

\begin{equation}\label{eq:a_Fe}
a_{\rm Fe}^{\rm met} = \gamma_{\rm Fe}^{\rm met} X_{\rm Fe}^{\rm met}
\end{equation}

\begin{equation}\label{eq:a_FeO}
a_{\rm FeO}^{\rm sil} = \gamma_{\rm FeO}^{\rm sil} X_{\rm FeO}^{\rm sil}.
\end{equation}

$a_{\rm FeO}^{\rm sil}$ and $a_{\rm Fe}^{\rm met}$ are the activities of FeO and Fe in the silicate melt (sil) and the metal (met), respectively. The $\gamma$'s are the activity coefficients. For non-ideal mixtures they are obtained from the asymmetric Margules equations using the fit parameters presented in \cite{Frost2010}. To account for ideal mixing the Margules parameters are smoothed above 50 GPa to drive the activity coefficients to unity at higher pressures (\cite{Schaefer2017}). Including the dependence on temperature, pressure and composition of the activity coefficients has no significant impact on the resulting estimates for the oxygen fugacity (e.g., \cite{Fischer2015}, \cite{Siebert2012}).


\begin{table}[t]
  \begin{center}
    \caption{Summary of the parameters considered in this study to characterize the interior structure and composition of Venus.}
    
    \begin{tabular}{lcl}
      \hline
       Parameter & Unit & Description \\
      \hline
         $\rm Mg\#$ & &  Total magnesium number [Mg]/[Fe + Mg]\\
         $\rm Si\#$ & & Total silicon number [Si]/[Si + Mg]  \\
         $P_{\rm CS}$ & GPa & Core segregation pressure \\
         $T_{\rm TBL}$ & K &  Temperature at the TBL\\
         $P_{\rm S}$ & GPa & Surface pressure \\
         $\Delta T_{\rm MTZ}$ & K & Temperature difference across the MTZ (fixed at 0 K)\\
         $\Delta T_{\rm CMB}$ & K & Temperature difference across the CMB \\
         $X_{\rm FeS}^{\rm Core}$ &  & Mole fraction of FeS in core [FeS]/[Fe + FeO + FeS + FeSi]\\
         $X_{\rm FeSi}^{\rm Core}$ &  &  Mole fraction of FeSi in core [FeSi]/[Fe + FeO + FeS + FeSi]\\
         $X_{\rm FeO}^{\rm Core}$  &  & Mole fraction of FeO in core [FeO]/[Fe + FeO + FeS + FeSi]\\
         $X_{\rm Fe}^{\rm Core}$  &  & Mole fraction of Fe in core [Fe]/[Fe + S + Si + O] \\
         $X_{\rm S}^{\rm Core}$ &  & Mole fraction of S in core [S]/[Fe + S + Si + O] \\
         $w_{\rm S}$ &  & Weight fraction of S in core \\
         $w_{\rm Si}$ &  &  Weight fraction of Si in core \\
         $w_{\rm O}$ &  & Weight fraction of O in core \\
         $X_{\rm FeO}^ {\rm Mantle}$ &  & Mole fraction of FeO in mantle [FeO]/[FeO + MgO + $\rm SiO_2$]\\
         $X_{\rm SiO_2}^ {\rm Mantle}$ &  & Mole fraction of $\rm SiO_2$ in mantle [$\rm SiO_2$]/[FeO + MgO + $\rm SiO_2$]\\
         $T_{\rm CS}$ & K & Core segregation temperature \\
         $R_{\rm Core}$ & km & Core radius \\
         $R_{\rm ICB}$ & km & Inner-to-outer core boundary \\
         $C/MR^2$ & & Normalized moment of inertia (referred to as MoI) \\
         $P_{\rm C}$ & GPa & Central pressure \\
         $T_{\rm C}$ & K &  Central temperature \\
         $f_{\rm O_2}$ &   & Oxygen fugacity of the mantle \\
    \hline

    \end{tabular}
    \label{tab:parameter_overview}
  \end{center}
\end{table}

\begin{table}[t]
  \begin{center}
    \caption{Summary of the different cases for the composition referred to in this study. Details on the Earth reference models can be found in Appendix \ref{app:earth_reference}. The bulk composition is defined via $\rm Mg\#$ and $\rm Si\#$.}
    
    \begin{tabular}{ll}
      \hline
       Composition & Description \\
      \hline
         Nominal &  Bulk composition and S content in the core from the Earth reference models.\\
         S-free & Bulk composition from the Earth reference models but no S in the core.\\
         S-rich &  Bulk composition from the Earth reference models with 20-50 mol\% FeS in the core. \\

    \hline

    \end{tabular}
    \label{tab:models_overview}
  \end{center}
\end{table}

\section{Results and discussion}\label{sec:results}
\subsection{Venus models}\label{sec:possible_models}

Venus' MoI is estimated to be 0.337 $\pm$ 0.024 (\cite{Margot2021a}). Our analysis considers all models for which Venus' radius matches within $\delta_R =$ 1\%. Three different cases for the core composition (S-free, nominal, S-rich) defined via the S content were considered (see section~\ref{sec:bulk_comp} for details). For these compositions three different cases are considered for the MoI: Models within the entire allowed range of 0.337 $\pm$ 0.024 (All MoI), models within 1\% of the lowest possible value (Low MoI), and models within 1\% of the highest possible value (High MoI). We find that not all values within 0.337 $\pm$ 0.024 are possible. As a result,  the highest and lowest MoI values are different for the various assumed compositions. The effects of turning off the core segregation model from section~\ref{sec:core_seg_icb} on the main results are discussed in Appendix~\ref{app:CS_model}.

\subsubsection{Internal profiles}\label{sec:profiles}

Figures~\ref{fig:venus_profiles_sulfur_poor} to \ref{fig:venus_profiles_sulfur_rich} show the possible temperature, pressure, and density profiles of the different ranges of MoI values and for the three compositions as defined above. The first row depicts all possible profiles for which the MoI lies in the allowed range. The second and third rows only show the profiles that lie within 1\% of the lowest and largest possible value for the MoI. The MoI is indicated in colors and the density profile for the Earth from the Preliminary Earth Reference Model (PREM) is shown for comparison. Furthermore, the resulting ranges for the core size and the conditions at the CMB are indicated by the shaded regions.

The most prevalent trend that emerges for the MoI in all cases are an overall increase with the mantle density. This behaviour is expected as higher mantle densities lead to a smaller density contrast between the mantle and the core. Increasing the size of the core can lead to both an increase and decrease of the MoI. The density of the mantle minerals is mainly influenced by the Fe content of the silicates. It is quantified by the amount of FeO, and hence related to the oxidation state of the mantle. The core size strongly depends on its composition. For a given Mg\#, larger amounts of lighter elements in the core lead to larger cores. Therefore, the MoI indirectly correlates with the core composition, and hence, the pressure gradient within the core. This typically leads to lower (higher) core pressures for high (low) MoI values.

The presence of a solid inner core and even a fully crystallized core cannot be excluded for any of the considered compositions. The highest possible pressures in the core are achieved for low MoI values while no obvious trends are  observed for the central temperature $T_{\rm C}$ in most cases (Figures~\ref{fig:venus_profiles_sulfur_poor}-\ref{fig:venus_profiles_sulfur_rich}). The ICB is dictated by the melting curve of Fe-alloys which shifts to lower temperatures with increasing S content (see Section \ref{sec:core_seg_icb}). A completely molten core without intrinsic heat sources cannot sustain thermal convection over long periods of time (\cite{Stevenson1983a}). In this case, the temperature profile within Venus' core could have become subadiabatic $\sim$1.5 Gyr ago.

\subsubsection{Internal structure and chemical composition}\label{sec:chem_comp}

Figures~\ref{fig:venus_all_scenarios} provide a schematic overview of the main composition and structure parameters. The left, middle, and right column correspond to the S-free, nominal, and S-rich models. The top, middle, and bottom rows correspond to all values, the lowest values, and the highest values of the MoI. The pressures and temperatures given for each layer transition correspond to the values in the lower of the two layers at the boundary.

For a S-free composition, the core contains up to $\sim$0.73 wt\% O and 13 wt\% Si. The amount of FeO in the mantle is poorly constrained between 0.05 and 7 mol\%. This leads to a large range of possible redox states of the mantle with $-\log{f_{\rm O_2}}=1.5-5.5$. If Venus is depleted in S, the radius of its core is predicted to be 2930-3610 km. The size of a solid inner core in this case would be 0-3330 km. The absence of S in the core leads to higher melting temperatures of the Fe-alloys. This allows for larger inner cores in comparison to the other compositions (see below). Finally, in the S-free case, the central pressure could be as high as 351 GPa, which is similar to the lower bound inferred for  reference models of the Earth (see Appendix \ref{app:earth_reference}).

If the true value of Venus' MoI is low, the possible ranges for the core size, the FeO content in the mantle, and the core composition would be much narrower. In particular, the core size would have to be between 3120 and 3610 km and contains no more than 0.39 wt\% O. Furthermore, the mantle could not contain more than 3.6~mol\% FeO. As expected, the FeO content in the mantle for the highest possible MoI can be larger and between 0.06 and 7~mol\%. In this case the density contrast between the core and the mantle is minimal. The core size is then predicted to be 2930-3510 km. An inner solid core of up to 3180 km could exist.
\\

The middle column of Fig.~\ref{fig:venus_all_scenarios} shows the results for the nominal composition. For these models a core composition with 0-1.3~wt\% O, 0.12--13~wt\% Si, and 3.9-7.9~wt\% S was obtained. This leads to a FeO content in the mantle of 0.03-9.8~mol\%. The size of the core in this case is  between 2960 and 3890 km. If an inner core is present, it could not be larger than 3300 km. The possible core- and mantle compositions strongly depend on the actual value of the MoI. 

 The second and third panels in the middle column in Fig.~\ref{fig:venus_all_scenarios} show the bracketing cases of the MoI with 0.317-0.323 and 0.331-0.338 for the nominal composition. For the lowest possible MoI the O content in the core is restricted to 0-0.42 wt\%, while the Si and S contents remain similar to the case that considers the entire range for the MoI. This results in 0.04-3.3~mol\% FeO in the mantle and hence a larger density contrast between the core and the mantle. The size of the core is 3120-3740 km. In this case the mantle of Venus is likely more reduced than the mantle of the Earth.

If the maximum MoI value for a nominal composition is assumed, the density contrast between the mantle and the core is smaller. This leads to much higher FeO contents in the mantle (4.3-9.8~mol\%) and an O content of 0.52-1.3~wt\% in the core. Furthermore, larger amounts of S in the core (4.3--7.8~wt\%) further decrease the density contrast by decreasing the core density. In this case the core size is 2980-3380 km and an inner core, if present, is limited to 2380 km.
\\

In the third column in Fig.~\ref{fig:venus_all_scenarios} the results for the S-rich models are shown. In this case the core could contain 0-2.9~wt\% O, 0.04-12~wt\% Si, and 9.1-22~wt\% S. This results in similar ranges for the MgO and $\rm SiO_2$ content of the mantle than for the S-free and nominal composition but a higher upper bound for the FeO content. This is due to the fact the O content of the cores of the S-rich models can be considerably larger than for the other compositions. It is possible that the presence of S influences the distribution of O and Si between the metal and the silicates (e.g. \cite{Fischer2015}). This could affect the correlations between the amount of S in the core and the composition of the mantle.

The core size of the S-rich models is 3010-4350 km. A core of Venus that is smaller than $\sim$3600 km would be consistent with a core depleted in S. A core size larger than about 3900 km could hint towards an enrichment in S with respect to the Earth reference models. The possible range for the MoI inferred for the S-rich models is 0.32-0.351. The upper bound for the MoI is considerably higher than for the other compositions while the lower bound is similar. This is expected as the main difference between these models is a larger amount of S in the core which decreases the density contrast between the mantle and the core.

The bracketing cases of the MoI for the S-rich models yield similar trends as for the other compositions. Lower MoI values generally require a larger density contrast between the core and the mantle. With this the FeO content in the mantles of the S-rich models with a low MoI cannot be higher than 3.9 mol\% and the maximum O content in the core is 0.56~wt\%. The possible range for the core size in this case lies between 3090 and 3920 km and a fully molten core is not possible. If the largest possible value for MoI is assumed, this range becomes 3060-3810 km. In our sample 677 models match the boundary conditions for Venus for an S-rich composition and the high MoI. Of these models only two were found for which an inner core could exist. An O content in the core of 1-2.9 wt\% and correspondingly a larger FeO content in the mantle, i.e. 5.5-15~mol\%, is required. Furthermore, the Si content of the core becomes very small for high MoI values.

\subsubsection{Mantle composition as a proxy for the core size}\label{sec:mantle_vs_core_comp}

 In this section the relation between the mantle composition and core size is investigated.  Fig.~\ref{fig:venus_mantle_comp_vs_core_size} presents the core radius as a function of the oxygen fugacity $\log f_{\rm O_2}$ (left) and the FeO content in the mantle $X_{\rm FeO}^{\rm Mantle}$ (right) for the different compositions. The MoI value  is indicated by different colors. The grey shaded region denotes the allowed values for the oxygen fugacity and  FeO content of the Earth reference models (see Appendix \ref{app:earth_reference} for details).
 
Different amounts of S primarily lead to lower core densities and larger cores at otherwise equal properties. Since potential effects of S on the metal-silicate partitioning of O and Si are not included, the mantle composition is not very sensitive to the S content. Larger amounts of S in the core lead to larger upper bounds of the O content. It is possible that additional trends for the different compositions would arise if partitioning effects are considered.
 
 The major trends evident in Fig.~\ref{fig:venus_mantle_comp_vs_core_size} are the negative correlation between the FeO content in the mantle and the core size and a decrease of the MoI value for smaller FeO contents. Both effects are readily explained by the fact that higher amounts of FeO in the mantle increase the mantle density and reduce the fraction of Fe in the core at a fixed bulk composition. This correlation between the mantle composition and the core size holds despite the large variability of the core composition.
 If the FeO content in Venus' mantle is similar to that of the Earth reference models and the nominal composition is assumed, its core radius is expected to be $\sim$2900-3600 km. Including the S-rich and S-free models this range becomes $\sim$2900-4000 km. This estimate does not depend on any knowledge of the S content in the core or further constraints on the MoI value. The mantle composition could, in principle, be further constrained from remote surface surveys and surface samples.

\subsubsection{Venus versus Earth}\label{sec:venus_vs_earth}

Fig.~\ref{fig:Earth_vs_Venus} compares the main parameters for the S-free, nominal, and S-rich composition to the Earth reference models. The composition of the core and mantle of Earth are taken from our model while the layer boundaries are taken from the PREM data (rounded to three significant digits). The MoI value is the measured value for the Earth. The relative sizes of the individual layers are to-scale and the possible ranges for the Venus models are indicated by dashed segments. The red segments denote lower bounds for the core size while the dark grey segments correspond to the lower bounds of the inner core.

Here, we focus on possible structural and compositional differences between Earth and Venus that could explain their measured properties. All the models presented in this study match the mean density of Venus by construction and hence incorporate the 2\% density deficit with respect to Earth. None of the inferred core or mantle compositions of the Earth can be excluded  for Venus. Therefore, Venus' mean density could be explained with a very similar bulk and core composition to that of Earth. 

Given the sparse data available for Venus it is also possible that the interior composition and element distribution significantly differ from those of Earth. 
For example, Venus' mantle could be more reduced than the mantle of the Earth. This scenario could possibly be excluded if a high MoI value is measured (see Fig.~\ref{fig:venus_all_scenarios}). In this case our model predicts that the redox state of the mantle could be similar to that of Earth, and possibly more oxidized. For the S-free models a high MoI would also be consistent with a more reduced mantle. The O content in Earth's core derived from our model is below 1 wt\% while for Venus up to 2.9 wt\% is predicted. The size of Venus' core cannot be constrained well from the models. This is due to the large variability in the possible core compositions and the element distribution between the core and mantle. 

If more stringent constraints on the abundances of the major oxides FeO, MgO and $\rm SiO_2$ were imposed it could be possible to narrow down the permissible range for the core size, even without more accurate MoI measurements or seismic data. In particular, for the nominal models a similar FeO content in the mantle than for the Earth would indicate that Venus' core could be similar in size to that of Earth (see Fig.~\ref{fig:venus_mantle_comp_vs_core_size}). The Si content in the cores of the nominal models could be much larger than is inferred for the Earth. If Venus has a more oxidized mantle than the Earth, its core must be considerably smaller for the nominal composition, but could be larger for S-rich compositions. For the S-free models Venus' core would be smaller than  Earth's core within the entire range of the FeO content in the mantle of Earth reference models.

\subsection{Comparison to previous studies}\label{sec:comp_prev_stud}

The models for the lowest possible MoI value for a S-free or nominal composition resemble the scenario from \cite{Lewis1972}. In this scenario a low oxidation state of the mantle and a core that was depleted in S relative to the Earth was proposed to explain Venus' density. The models with S and the highest possible value of the MoI are akin to that of \cite{Ringwood1977}. In this case the oxidation state of the mantle would be higher. It was inferred that Venus would currently contain about 13.4 wt\% FeO in the mantle or [FeO]/[FeO + MgO] = 0.24 and $\sim$16 mol\% S in the core (\cite{Ringwood1977}). This would lead to a core mass fraction of $\sim$0.23 in comparison to 0.32 for Earth. These predictions are within or close to the inferred ranges from our models for the highest MoI of the S-rich models. In particular, for these models we predict a core mass fraction of 0.22-0.38, a S content in the core of 21-32~mol\%, and [FeO]/[FeO + MgO] = 0.11-0.28. The oxygen fugacity is $-\log{f_{\rm O_2}} = 0.7-2.1$ which is more oxidized than the Earth reference models which have $-\log{f_{\rm O_2}} = 1.8-2.8$ (see Appendix~\ref{app:earth_reference}).

Models with lower oxygen fugacities tend to result in a lower MoI due to the larger mass concentration in the core. It seems therefore likely that the different scenarios proposed by \cite{Lewis1972} and \cite{Ringwood1977} are quantitatively represented by the bracketing cases for the MoI found in this study. The idea that reliable measurements of the MoI could allow to distinguish between these two (and other) scenarios has been entertained early on (e.g. \cite{Ringwood1977}, \cite{Goettel1981}). 
If Venus' MoI value is measured to be at the high end of the possible values, this could be strongly in favor of the scenario proposed by \cite{Ringwood1977}.

Previous studies used the density profile of the Earth to construct a scaled model for the density as a function of depth for Venus (e.g. \cite{Aitta2012}, \cite{Steinberger2010}). \cite{Dumoulin2017} presented a series of different interior models for Venus and computed the core radius and MoI for the different cases. They employed four different composition models which were based on assumptions regarding the composition of the solar nebula and the accretion history. Furthermore, in this study only two end-member temperature profiles (referred to by the authors as "hot" and "cold") were considered. These profiles were inferred from assumptions about the similarity of Earth and Venus in combination with recent data on topography and gravity anomalies (\cite{Steinberger2010}, \cite{Armann2012}). Amongst their models the largest variation of the moment of inertia was found to be no more than about 5\%, which is too low to be resolved within the current observational uncertainty.

The central temperature and pressure of Venus were estimated to be 5200 K and 275 GPa (\cite{Aitta2012}). \cite{Steinberger2010} obtained a central pressure of about 285 GPa. A more recent estimate  predicted a range of $271-299$ GPa (\cite{Dumoulin2017}). The latter was obtained from previously published temperature profiles (\cite{Steinberger2010}, \cite{Armann2012}). Furthermore, the core radius of Venus was estimated to be $\sim$3186 km by \cite{Steinberger2010} and 3228 km by \cite{Aitta2012}. \cite{Dumoulin2017} obtained values between 2941 km and 3425 km. These estimates are well within the possible ranges obtained here. Our lower limit for the core size differs from that obtained by \cite{Dumoulin2017} only by 10-70 km for the three compositions considered here. However, our model predicts a much larger upper limit for the core size. This is because large amounts of lighter elements were allowed to be present in the core and more variable mantle compositions were considered here. For the nominal composition and an FeO content in the mantle similar than in Earth's mantle, the core size is estimated to be $\sim$2900-3600 km. This range is only slightly larger than that inferred by \cite{Dumoulin2017}.

The MoI is estimated to be 0.317-0.33 for a S-free, 0.317-0.338 for a nominal and 0.32-0.351 for a S-rich composition. These ranges are somewhat larger, but compatible with previous estimates of 0.327-0.342 and 0.329-0.341 (\cite{Dumoulin2017}, \cite{Zhang1995}). 
Fig.~\ref{fig:MoI_ranges} shows the predicted ranges for MoI from various authors including the results obtained here for the different assumed compositions. The shaded region depicts the currently allowed range from observations (\cite{Margot2021a}) while the black bars show the inferred ranges from different models. The MoI of Earth, Mars, Mercury, and the Moon are shown for reference. The models from \cite{Zhang1995}, \cite{Aitta2012} and \cite{Dumoulin2017} assumed a composition similar to the Earth and obtained the density profile in the mantle from scaling the density of the Earth. This leads to a narrower range for the MoI. Although the model approaches in these studies are different the inferred ranges are similar. This illustrates the redundancy in Venus structure models and confirms that an accurate MoI measurement alone, although highly desirable, would not strongly constrain Venus' internal structure.

\section{Summary}\label{sec:summary}
This study presents interior structure models for Venus for different bulk compositions. The mantle composition was obtained from chemical equilibrium calculations using published metal-silicate partitioning data. Core-mantle equilibration was assumed to have occurred as a single-stage process. We further assume that the pressure range over which core segregation occurred in Venus was similar to that of Earth. This model was employed to infer possible interior models for Venus and assesses potential structural and compositional differences between Venus and Earth.

Three different compositions, S-free, nominal, and S-rich, were considered. While we can impose constraints on some of the composition and internal structure parameters from available data, considerable redundancies in possible structure models for Venus remain. A better understanding of the influence of S and other elements on the metal-silicate partitioning of O and Si could help to mitigate these redundancies. Furthermore, constraining the composition and chemical state of Venus' mantle could provide tighter constraints on possible interior models and in particular on the core size. Our key results can be summarized as follows:

\begin{enumerate} 
    \item In the framework of our model, higher concentrations of S in the core correspond to higher concentrations of O in the core while the range for the Si concentration remains unchanged. This result may need modification if partitioning of Si and O into the core is affected by the presence of S in the metal.
    
    \item Venus could be much more oxidized than Earth. As a consequence the FeO content in Venus' mantle could be significantly higher than that of Earth. High values of the MoI would be in favor of models with high oxygen fugacities. Low values of the MoI could indicate a more reduced state.
    \item Currently available constraints cannot rule out the absence of a solid inner core.

    \item Modelling the influence of S and other elements on the metal-silicate partitioning of O and Si could considerably mitigate the redundancy of possible composition and differentiation models for Venus. Experiments to quantify these effects are highly desirable.

    \item The core size of the Venus models from all considered compositions is estimated to be 2930-4350 km. If the S content in the core and the FeO content of the mantle are similar to those inferred from the Earth reference models, the permissible range for the core size is approximately 2900-3600 km.

    \item  The pressure and temperature conditions at the center of Venus could be similar to those inferred for Earth.
    
    \item The inferred ranges for the MoI depend on the assumed bulk compositions and are found to be 0.317-0.33, 0.317-0.338, and 0.32-0.351 for a S-free, a nominal, and a S-rich composition, respectively. These ranges are in good agreement with previous studies.
    
\end{enumerate}

Currently, there are two future space missions dedicated for the exploration of Venus. NASA has recently announced the selection of DAVINCI (Deep Atmosphere Venus Investigation of Noble gases, Chemistry, and Imaging) and VERITAS (Venus Emissivity, Radio Science, InSAR, Topography, and Spectroscopy) for further studies, with a launching date by the end of this decade. EnVision is a planned low altitude orbiter and currently a candidate for ESA's cosmic vision program. 
Numerical simulations suggest that it could reach a precision for the tidal Love number and the moment of inertia of Venus of 0.3\% and 1.4\%, respectively (\cite{Rosenblatt2021}). Although, EnVision is expected to be launched after the mid 2030's, it will constrain the composition and structure of our twin planet. This could promote our understanding of planet formation mechanisms and shed further light into why Venus and Earth have evolved so differently despite their similarities in size, mass, and heliocentric distance.

In preparation for such missions theoretical predictions of possible compositions and internal structures are crucial for interpreting the acquired data. The model presented here could be used to guide such interpretations in the near future. 
Accurate measurements of Venus' gravity field and further constraints on its mantle composition would narrow down the possibilities for its core size and composition as well as its thermal state. In combination with a more complex chemical and thermal model for core segregation, including the effects on metal-silicate partitioning of S and other elements, it could be possible to constrain the chemical composition and state of the core of Venus even in the absence of seismic data.


\section*{Acknowledgement}
We thank Dave Stevenson and Morris Podolak for discussions. 
This work has been carried out within the framework of the National Centre of Competence in Research (NCCR) PlanetS supported by the Swiss National Science Foundation (SNSF).

\bibliography{library}{}
\bibliographystyle{aasjournal}

\newpage
\appendix

\section{The effect of the core segregation model}\label{app:CS_model}

\begin{figure}[ht]
\includegraphics[width=\textwidth]{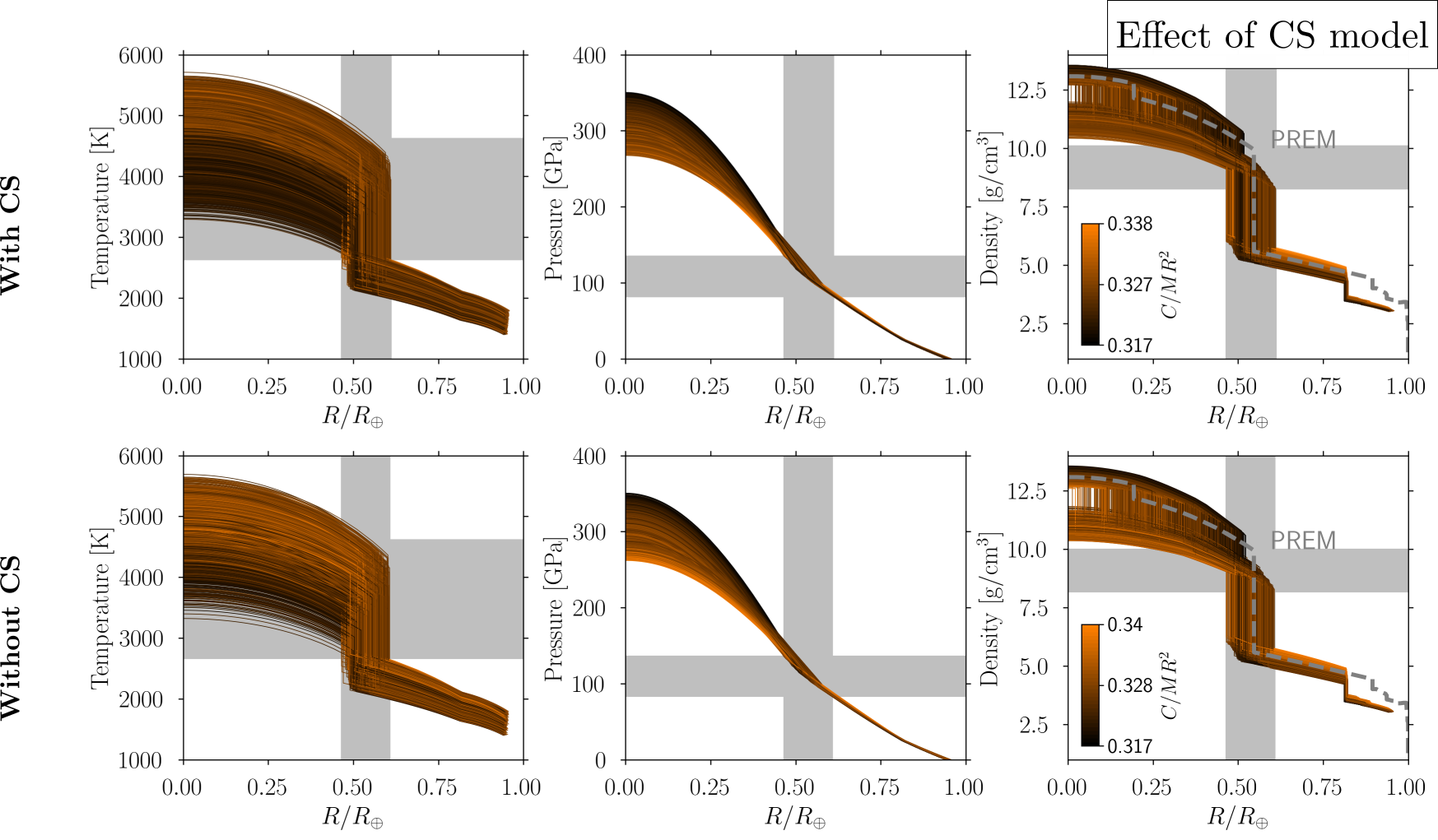}\caption{Effect of the core segregation model on the interior profiles for the nominal composition. The top panels are equivalent to the top panels in Fig.~\ref{fig:venus_profiles_earth_like} and the bottom panels correspond to the results obtained without invoking the core segregation model.}
\label{fig:CS_vs_no_CS_profiles}
\centering
\end{figure}

\begin{figure}[ht]
\includegraphics[width=\textwidth]{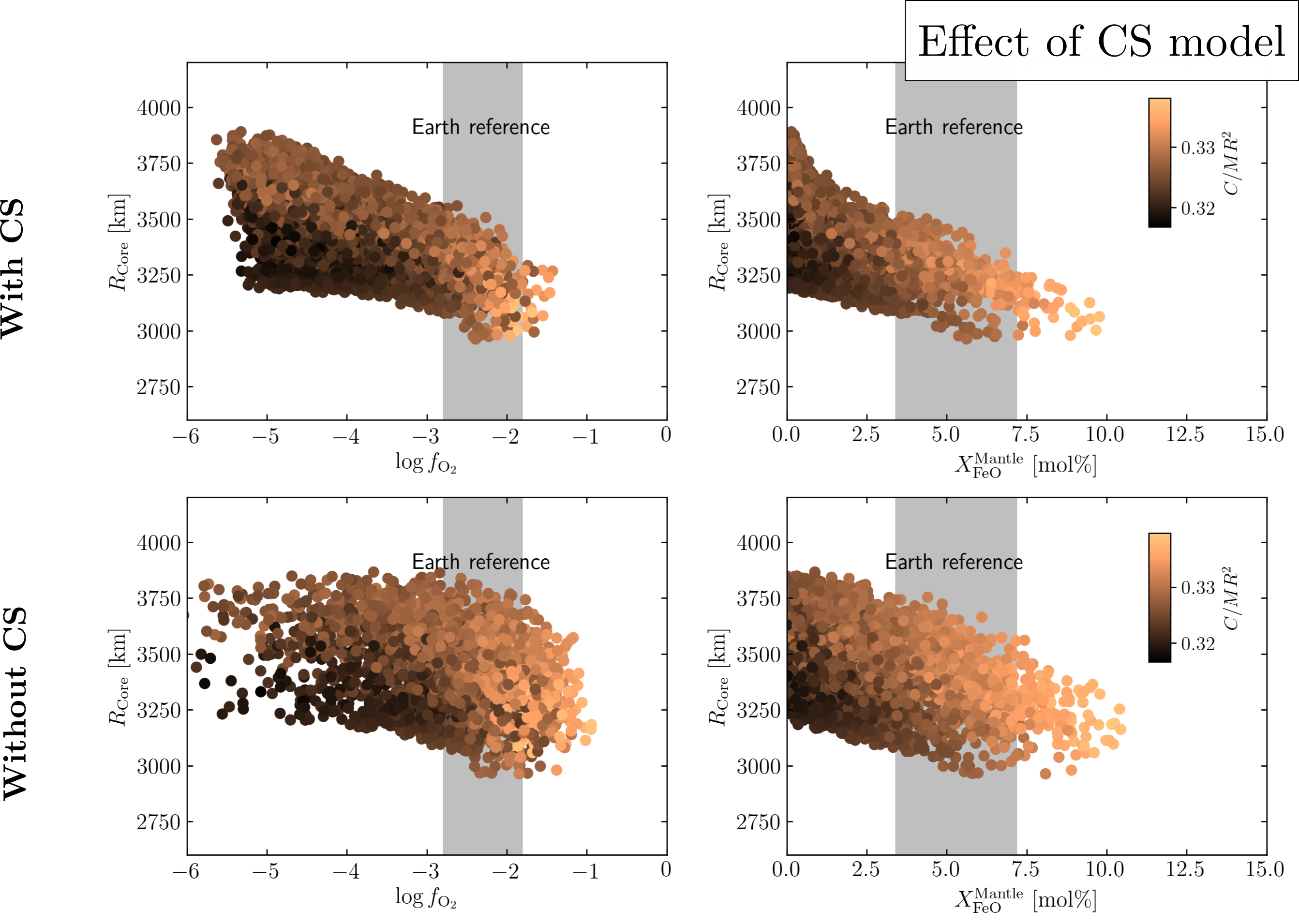}\caption{Effect of the core segregation model on the correlation between the mantle composition and the inferred core size of Venus for the nominal composition. The top panels are equivalent to the top panels Fig.~\ref{fig:venus_mantle_comp_vs_core_size} and the bottom panels correspond to the results obtained without invoking the core segregation model.}
\label{fig:CS_vs_no_CS_correlation}
\centering
\end{figure}

Here the sensitivity of our results to the model for chemical equilibration during core segregation is assessed. To this end an additional set of Venus models probing the same parameter ranges as given in Table~\ref{tab:input_parameter_ranges} was created without invoking the core segregation model from section~\ref{sec:core_seg_icb}. For these models the mantle composition is not constrained from the core composition. To illustrate the effects on the main results a comparison of the two cases is given for the nominal composition ($X_{\rm FeS}^{\rm Core}=0.08-0.15$). Fig.~\ref{fig:CS_vs_no_CS_profiles} shows the interior profiles for the nominal models using the core segregation model (top panels) and without the core segregation model (lower panels). Relaxing the constraints on the mantle composition has little effect on the overall range of possible interior structures. Fig.~\ref{fig:CS_vs_no_CS_correlation} shows a comparison of the correlation between the mantle composition and the core size. The spread of permissible core sizes at a given mantle composition is enhanced if the core segregation model is turned off. A comparison of the permissible ranges for a selection of parameters is given in Table~\ref{tab:parameter_study}. The first column shows the inferred ranges for the nominal bulk composition as presented in the main text. The second column corresponds to the ranges obtained when the core segregation model is turned off and the third column considers a wider range for the core segregation pressure (see Appendix \ref{app:effect_P_CS}). The most prominent effect of the CS model is observed for the O content in the core and the oxygen fugacity of the mantle. This is not surprising as the O and Si content in the core are controlled by the core segregation process.

\begin{figure}[ht]
\includegraphics[width=\textwidth]{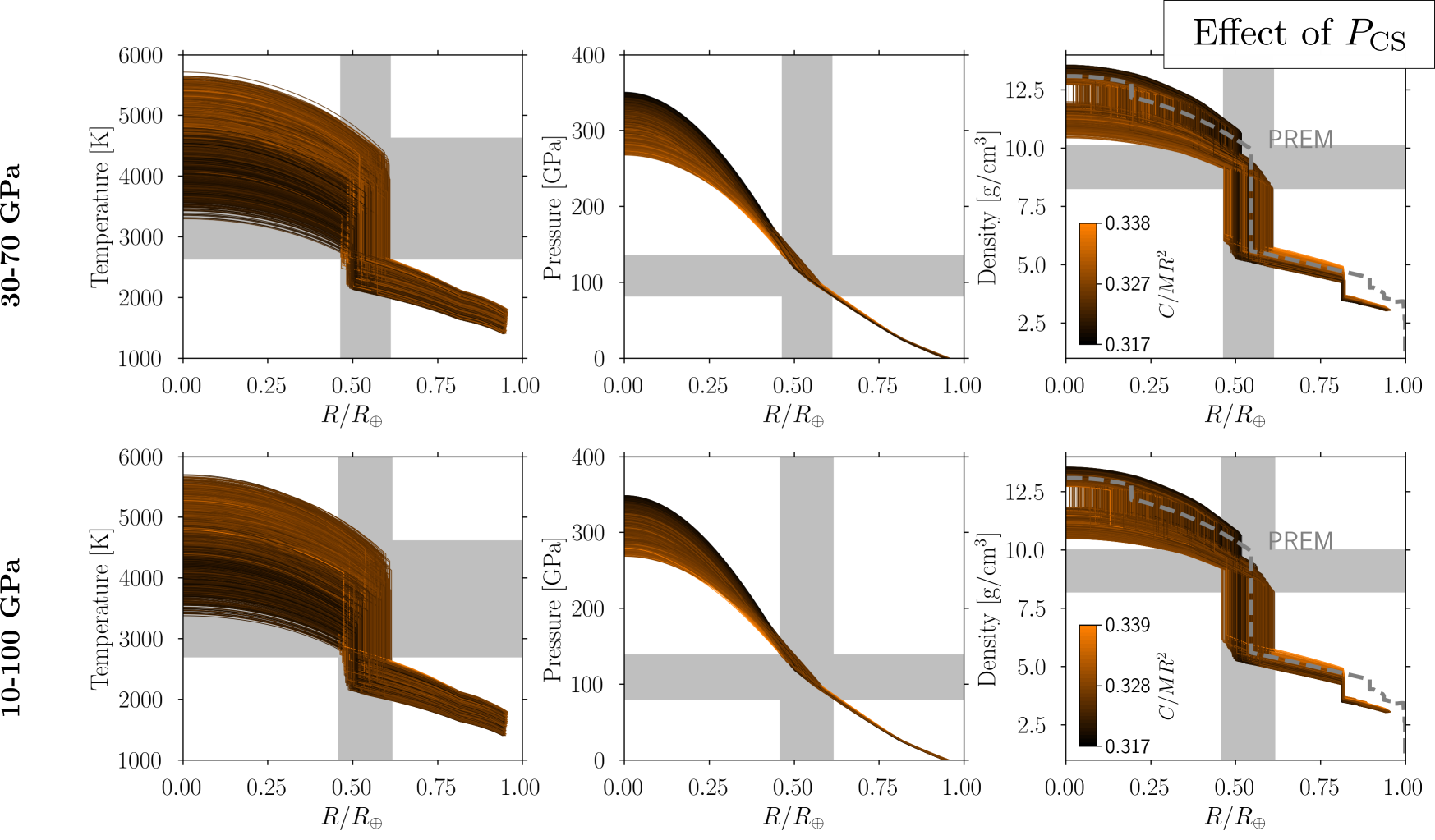}\caption{Effect of the range of the core segregation pressure on the interior profiles for the nominal composition. The top panels are equivalent to the top panels Fig.~\ref{fig:venus_profiles_earth_like} and the bottom panels correspond to the results obtained with $P_{\rm CS}=10-100$ GPa.
\label{fig:wide_profiles}}
\centering
\end{figure}

\begin{figure}[ht]
\includegraphics[width=\textwidth]{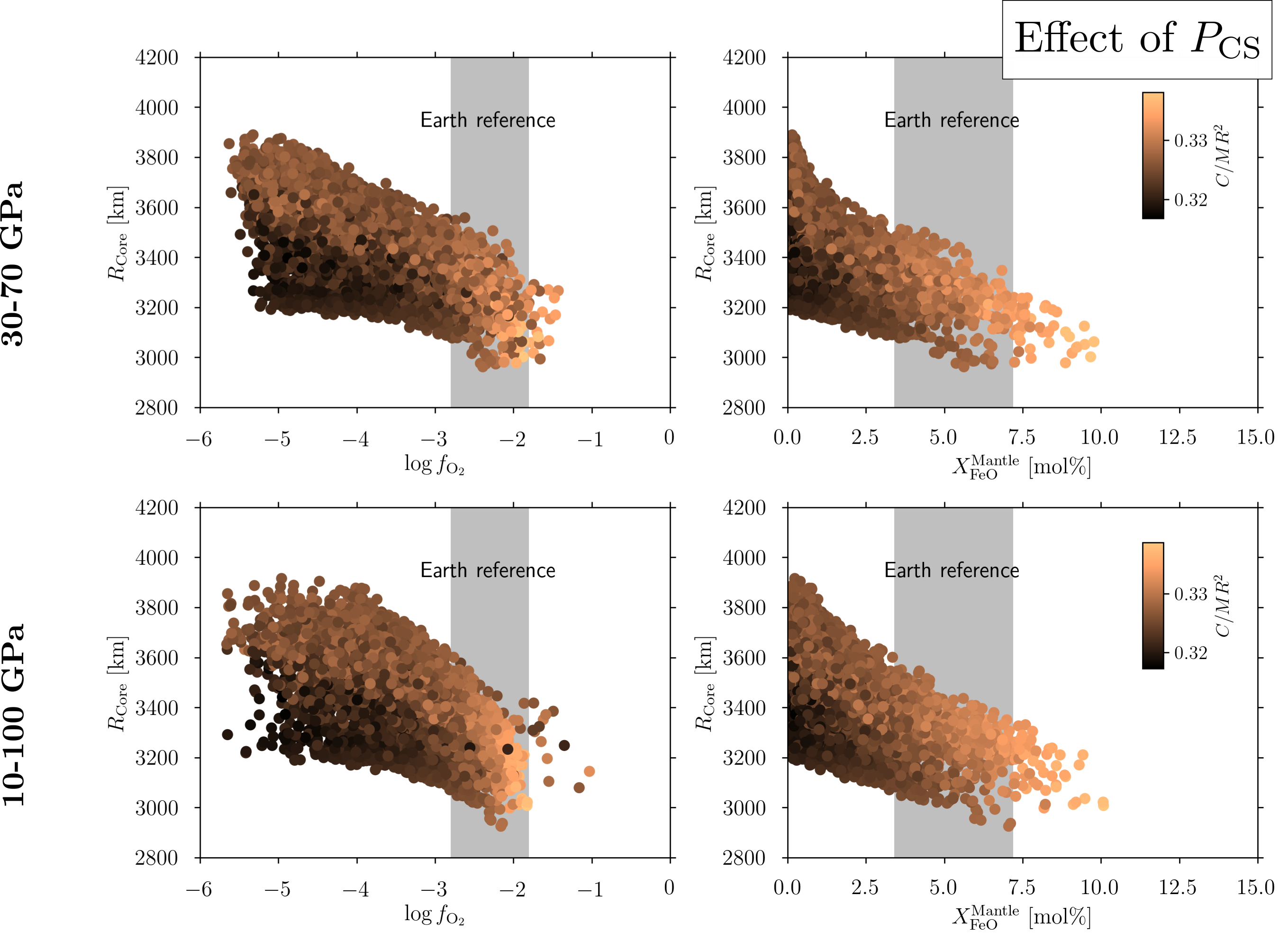}\caption{Effect of the range of the core segregation pressure on the correlation between the mantle composition and the inferred core size of Venus for the nominal composition. The top panels are equivalent to the top panels Fig.~\ref{fig:venus_mantle_comp_vs_core_size} and the bottom panels correspond to the results obtained with $P_{\rm CS}=10-100$ GPa.}
\label{fig:wide_PCS_correlation}
\centering
\end{figure}

\section{The effect of the chosen range of the core segregation pressure}\label{app:effect_P_CS}

The Venus models presented in this study were constructed assuming the core segregation process to have been similar to that in the Earth. In this section the effects of relaxing this assumption on the main results are discussed. To this end a set of Venus models with the nominal bulk composition was created for a wider range of the core segregation pressure of $P_{\rm CS} = 10-100$ GPa. Pressures above 100 GPa were not probed by the experiments of \cite{Fischer2015}. A comparison of the resulting pressure, temperature, and density profiles with the models presented in the main text are shown in Fig.~\ref{fig:wide_profiles}. Fig.~\ref{fig:wide_PCS_correlation} shows the same comparison for the correlation between the mantle composition and the core size. The third column in Table~\ref{tab:parameter_study} shows the resulting ranges for some of the major parameters investigated in this study. Interestingly, allowing a much wider range for the core segregation pressure has only minor effects on most of the studied parameters. The only significant effect was observed for the O content in the core which can become larger if a wider pressure range is considered. The effect of turning off the core segregation model on the core composition and oxygen fugacity is larger than the effect of considerably increasing the probed range of the core segregation pressure. This means that the main results of this study are less sensitive to the chosen range of the core segregation pressure. It is therefore likely that using a more complex multi-stage core segregation model would only have minor effects on the main results. This should be investigated further.

\section{Thermal equations of state}\label{app:theos}

The EoS used in this study are identical to the ones used in a previous study (see Appendix B in \cite{Shah2021}). The EoS parameters for the materials that have been added to the model or for which changes were adopted are summarized in Table~\ref{tab:coeffs_EoS}. The parameters are given for the Mie-Grüneisen-Debye EoS (MGD) and the 3rd order Birch-Murnaghan EoS (BM3). The mean densities of mixtures were computed using a linear mixing law.

\begin{table}[h]
    \caption{ EoS parameters for the materials that have not been described in \cite{Shah2021} or for which modifications have been adopted.}
    \begin{center}
        \begin{tabular}{ cccccccccccc }
         \hline
         material & EoS & $\rho_0$ & $K_{T,0}$ & ${K^\prime}_{T,0}$  & $\gamma_0$ & $\theta_{D,0}$ & $q$ & $ a_T$ & $ b_T$ & $ c_T$ & $a_P$\\
          & & $\rm [ kg \ m^{-3}]$ & [GPa] &  [K] & & [K] &  & $\rm [K^{-1}]$ & $\rm [K^{-2}]$ & [K] & $\rm [GPa \ K^{-1}$ \\
         \hline
         FeS  & MGD & 4900$^{(a)}$ & 135$^{(a)}$ & 6$^{(a)}$ & 1.36$^{(a)}$ & 998$^{(a)}$ & 0.91$^{(a)}$ & ... & ... & ... & ...\\
         FeSi  & MGD & 6543$^{(b)}$ & 230$^{(b)}$ & 4.17$^{(b)}$ & 1.3$^{(b)}$ & 417$^{(b)}$ & 1.7$^{(b)}$ & ... & ... & ... & ...\\
         FeO  & MGD & 5862$^{(c)}$ & 137.8$^{(c)}$ & 4$^{(d)}$ & 1.45$^{(a)}$ & 430$^{(a)}$ & 3$^{(a)}$ & ... & ... & ... & ...\\
         MgO  & MGD & 3583$^{(d)}$ & 160.2$^{(d)}$ & 3.99$^{(d)}$ & 1.524$^{(d)}$ & 773$^{(d)}$ & 1.65$^{(d)}$ & ... & ... & ... & ...\\
         $\rm Mg_2 Si_2 O_6^{(a)}$ & BM3 & 3215 & 111 & 7 & ... & ... & ... & 2.86e-5& 7.2e-9 & 0 & 0 \\
         $\rm Fe_2 Si_2 O_6^{(a)}$ & BM3 & 4014 & 111 & 7 & ... & ... & ... & 2.86e-5 & 7.2e-9 & 0 & 0 \\
         \hline
        \end{tabular}
        \label{tab:coeffs_EoS}
        
        {\raggedright \textbf{References.} (a)~\cite{Alibert2014}, (b)~\cite{Fischer2015}, (c)~\cite{Fischer2011}, (d)~\cite{Dorogokupets2010} \par}
    \end{center}
\end{table}

\section{Earth reference models}\label{app:earth_reference}
\begin{figure}[t]
\includegraphics[width=\textwidth]{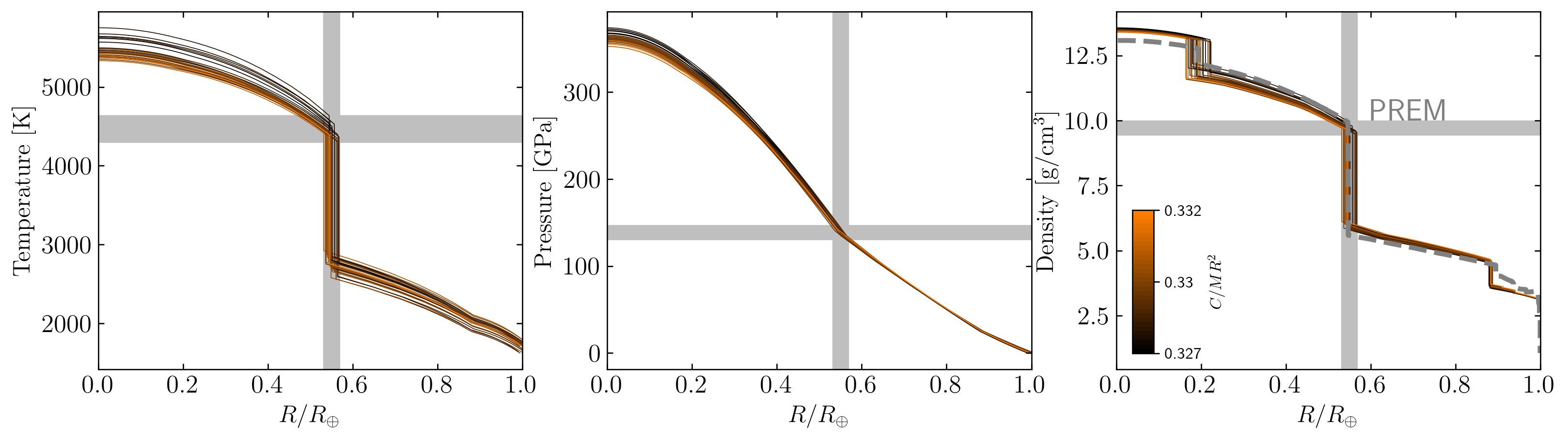}\caption{Temperature (left), pressure (center), and density (right) profiles of the Earth reference models. The PREM data (grey dashed curve) is shown as a reference. The density in the inner core is somewhat overestimated as it was assumed to consist of pure iron. In order to match the mean density and the MoI of the Earth, this results in a underestimation of the density in the outer core.}
\label{fig:earth_profiles_anhydrous}
\centering
\end{figure}

In order to validate our model and to define the nominal bulk composition used in this study the model was applied to the interior of modern-day Earth. For this additional constraints provided by the PREM data were imposed on the interior structure. In particular, we require the ICB, the CMB, and the MTZ to be matched with no more than 200 km deviation, respectively. Furthermore, the densities at these locations were required to agree with the PREM data within 5\%. The model uncertainty on the radius and MoI of Earth are set to be 1\%. It is commonly acknowledged that pure iron overestimates the density of the inner core of the Earth by $\sim$3\%. This hints towards the presence of some lighter elements in the inner core. Here, the inner core was modelled as pure iron for simplicity. This results in a slight overestimation of the inner core density and a slight underestimation of the outer core density in our models. Our models furthermore somewhat overestimate the density of the lowermost mantle and underestimate the density in the MTZ. 
This could be explained by the finding that the MTZ of Earth could be enriched in Fe with respect to the rest of the mantle (\cite{Liu2020}, \cite{Zhu2019}). This possibility was not considered here. The ranges for the probed parameters are listed in Table \ref{tab:input_parameter_ranges_earth}. A total of 82,821 models were created within this ranges, 28 of which satisfied the above conditions. The resulting permissible ranges for all relevant input and output parameters are summarized in Tables \ref{tab:input_parameters_earth} and \ref{tab:output_parameters_earth}. The temperature, pressure, and density profiles as a function of the normalized radius are shown in Fig.~\ref{fig:earth_profiles_anhydrous}.

 Ranges for the central pressure and central temperature of 352-374 GPa and 5330-5750 K are obtained. The FeO and $\rm SiO_2$ contents in the mantle are $\sim 3.4-7.2\%$ and $\sim 44-50\%$. These ranges are somewhat wider but in good agreement with estimates of the composition of basaltic rocks on Earth. The bulk Mg and Si contents are $\rm Mg\#=0.48-53$ and $\rm Si\# = 0.47 - 0.52$. The outer core is estimated to contain 4.6-7.6 wt\% S, 0.21-1.3 wt\% Si, and 0.37-0.95 wt\% O. The estimates for O and Si are consistent with recently published values (\cite{Hirose2021}) of 0-4 wt\% Si and 0.8-5.3 wt\% O. The S content in our model is larger than the 1.7 wt\% obtained by these authors. This is due to the fact that the presence of Ni, C, and H in the core was not accounted for in this study. \cite{Hirose2021} estimated a total of $\sim 5.5$ wt\% contained in these elements. Replacing some of the S with Ni, C, and H in our model (neglecting the effect on the outer core density) in order to match the Ni, C, and H content from \cite{Hirose2021} would lead to $\lesssim 2$ wt\% S in the outer core. This is consistent with their estimate for the S content. These results show that our model is capable of reproducing the composition of the Earth but with the constraints employed in this study the inferred allowed ranges for some parameters are larger than those obtained for the Earth from measurements.
 
 In addition to the presence of Ni, C, and H, the partitioning of lighter elements between the inner and the outer core was not considered in this study and should be investigated further. The total S content corresponds to a molar FeS content of roughly 8.6-15\%. The inferred composition of the Earth reference models were used to constrain the compositions of the Venus models presented in section~\ref{sec:results}. The allowed ranges of $\rm Si\#=0.47-0.56$ and $\rm Mg\#0.47-0.53$ were taken from the bracketing cases of \cite{Sotin2007} for all Venus models (see Tables~\ref{tab:input_parameters_earth} and \ref{tab:output_parameters_earth}). The FeS content of the core for the nominal composition was constrained from the Earth reference models to $X_{\rm FeS}^{\rm Core}=0.08-0.15$.

It is interesting to note that the possible range for the core segregation pressure spans the total range of  30 - 70 GPa but the chemical composition of both the core and the mantle are rather restricted. This is a desirable outcome for potential applications of our model to exoplanets. If the results were very sensitive to the core segregation pressure the model would need to be constrained to a narrower range. This would be challenging for exoplanets. The fact that the Earth can be reproduced within a rather wide pressure range indicate that only a very crude estimation of $P_{\rm CS}$ is required to characterize the composition of a planet (at least in terms of the major oxides FeO, MgO, and $\rm SiO_2$). On this level of precision a simple scaling law $P/P_\oplus \sim \left(M/M_\oplus\right)^\eta$ is likely sufficient to adopt adequate pressure ranges for planets over a wider mass range. With this the difference between the Earth and Venus in terms of the core segregation pressure is likely of minor importance because of their similar masses. The effect is expected to become relevant only if planets over a wider mass range, say 0.1 - 10 $M_\oplus$ for terrestrial exoplanets, is considered. It is expected that this would give rise to systematic trends in the composition and chemical states of exoplanets as a function of their total mass (see also \cite{Schaefer2017}). These trends would be based on the differences in chemical equilibrium between core and mantle due to the differences in the core segregation pressure. Such trends could assist in characterizing exoplanets and should be investigated further.

\section{Testing convergence}\label{app:convergence}
Our data set consists of a total of 98,046 individual models distributed over the parameter space summarized in Table \ref{tab:input_parameter_ranges}. From this we estimate the maximum allowed ranges for each parameter. However, due the finite number of models it has to be ensured that these ranges are representative and that they would not get significantly enlarged if the total number of models was increased. 

We therefore employ a simple test to confirm the convergence of the results. Random subsets of the data of variable sizes have been selected to reassess the resulting parameter ranges. The chosen sample sizes were: 

\begin{ceqn}\label{eq:N_sample}
\begin{align}
    N_j = N_{\rm tot}/2^{(1+j)},
\end{align}
\end{ceqn}
where $N_{\rm tot}$ denotes the total number of models and $j = 0,1,2,3,4,5$. For each chosen sample size we iterate this procedure $k=10,000$ times and  compute the root mean square deviation of the lower and upper bounds for each parameter $Q$:

\begin{ceqn}\label{eq:RMSD}
\begin{align}
     {\rm{RMSD}}_j(Q) = \sqrt{\frac{\sum_{i=1}^k \left(x_{ij}-\bar{x}_j\right)^2}{k-1}},
\end{align}
\end{ceqn}
where $x_{ij}$ is the inferred upper or lower bound for the sample $i$ of size $N_j$, $\bar{x}_j$ is the mean upper- or lower bound inferred for the sample size $N_j$ and $k$ is the number of samples for each sample size. The relative variability is then defined as:

\begin{ceqn}\label{eq:variability}
\begin{align}
     \sigma_j(Q) \equiv \frac{{\rm{RMSD}}_j(Q)}{{\bar{x}}_j}.
\end{align}
\end{ceqn}

\begin{figure}[t]
\includegraphics[width=\textwidth]{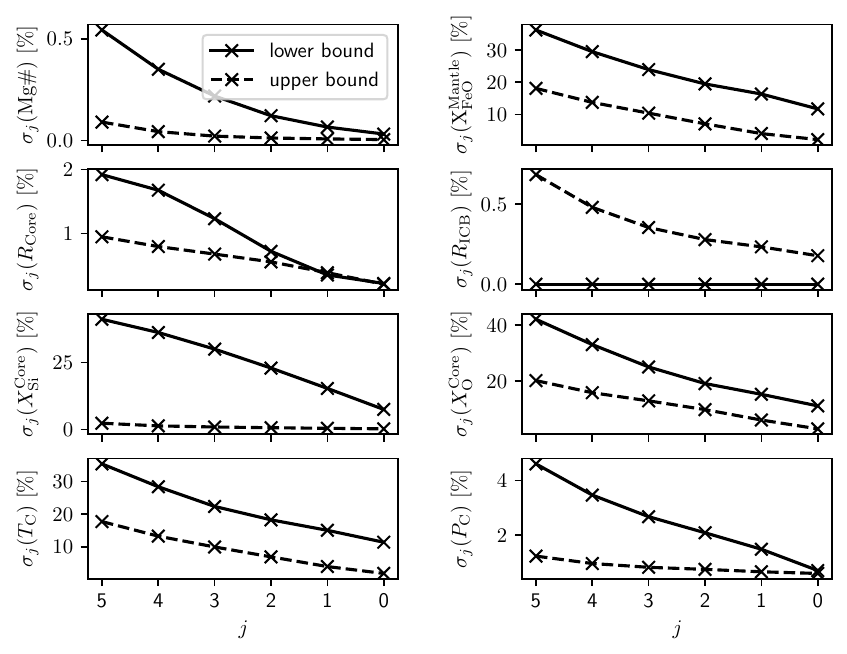}\caption{Variability in the ranges for the a selection of important parameters as a function of the size of the sample $N/2^{(1+n)}$ considered. The shown data corresponds to the models with a nominal composition.}
\label{fig:venus_conv}
\centering
\end{figure}

The resulting $\sigma_j(Q)$ for some of the tracked parameters are shown in Fig.~\ref{fig:venus_conv}. The solid and dashed curves correspond to the results for the lower and upper bounds on each parameter, respectively. We find that the variability monotonically decreases for all parameters. The $\sigma_{0}(Q)$ can therefore be viewed as a measure for the variability of the parameter ranges estimated in this study. For the cases where the only high or low MoI were considered the number of models that match the boundary conditions for Venus is smaller. The corresponding variability of the parameter ranges from this approach would be larger in these cases.

\begin{table}[ht]
  \begin{center}
    \caption{Comparison of the permissible ranges for some of the major parameters of the results presented in the main text (Main), cases where no core segregation model was used (No CS) and cases where a core segregation pressure of 10--100 GPa was used (Wide $P_{\rm CS}$).}
    \begin{tabular}{lcccc}
      \hline
       Parameter & Unit &  Main & No CS & Wide $P_{\rm CS}$  \\
      \hline
        $C/MR^2$ & & 0.317-338 & 0.317--0.34 & 0.317--0.339\\
         $X_{\rm FeO}^ {\rm Mantle}$ & mol\%  & 0.03-9.8 & 0--10 & 0--10 \\
         $X_{\rm SiO_2}^ {\rm Mantle}$ & mol\% & 38-50 & 38--50 & 38--50 \\
         $R_{\rm Core}$ & km & 2960-3890  & 2960--3870 & 2930--3920\\
         $R_{\rm ICB}$ & km & 0-3300 & 0--3330 & 0--3300\\
         $-\log(f_{\rm O_2})$ &   & 1.4-5.6 & 0.94-11 & 1--5.7  \\
         $w_{\rm S}$ & wt\% & 3.9-7.9 & 3.8-7.8 & 3.9--7.9 \\
         $w_{\rm Si}$ & wt\% & 0.12-13 & 0-12 & 0.06--13 \\
         $w_{\rm O}$ & wt\% & 0-1.3 & 0--2.6 & 0--1.7 \\
         $P_C$ & GPa & 267-350 & 262-351 & 268-348 \\
         $T_C$ & K & 3290-5710 & 3320-5690 & 3370--5700 \\
    \hline
    \end{tabular}
    \label{tab:parameter_study}
  \end{center}
\end{table}

\begin{table}[h]
  \begin{center}
    \caption{Probed ranges for input parameters for the Earth models.}
    \begin{tabular}{lccccccc}
      \hline
       Mg\#  & $P_{\rm CS}$ & $T_{\rm TBL}$ & $\Delta T_{\rm CMB}$ & $X_{\rm FeO}^{\rm Core}$ & $X_{\rm FeS}^{\rm Core}$ & $X_{\rm FeSi}^{\rm Core}$ \\
        & [GPa] & [K] & [K] & [K] & & & \\
      \hline
         0.47-0.53 & 30-70 & 1400-1800 & 500--1800 & 0-0.1 & 0-0.2 & 0-0.1\\
    \hline
    \end{tabular}
    \label{tab:input_parameter_ranges_earth}
    
  \end{center}
\end{table}

\begin{table}[t]
  \begin{center}
    \caption{Permissible ranges for the input parameters for the Earth reference models.}
    \begin{tabular}{lccc}
      \hline
       Parameter & Unit & Model range  & Literature  \\
      \hline
         $\rm Mg\#$ & & $0.48-0.53$ & $0.47-0.53$ $^{(a)\star}$ \\
         $P_{\rm CS}$ & GPa & 30-70 & 30-70 $^{(b,c,d,e,f,g)}$ \\
         $T_{\rm TBL}$ & K & 1620--1800  & 1600-1800
    $^{(h)}$ \\
         $\Delta T_{\rm CMB}$ & K & 1510-1800 & 500-1800 $^{(i,s)}$ \\
         $X_{\rm FeS}^{\rm Core}$ & mol\% & 8.6-15  & 3.6-36$^{(l,m,d,e,r,p,q,n,o,r)}$ \\
         $X_{\rm FeSi}^{\rm Core}$ & mol\%  & 0.46-2.7 & 2.1-27$^{(l,m,d,e,r,p,q,n,o,r)}$ \\
         $X_{\rm FeO}^{\rm Core}$ & mol\% & 1.4-3.6 & 1.8-12$^{(l,m,d,e,r,p,q,n,o,r)}$\\
         $w_{\rm S}$ & wt\% & 4.6-7.6 & 1.7-15 $^{(l,m,n,o,r,t)}$ \\
         $w_{\rm Si}$ & wt\% & 0.21-1.3 & 0-10 $^{(l,m,d,e,r,t)}$ \\ 
         $w_{\rm O}$ & wt\% & 0.37-0.95 & 0.5-5.3 $^{(l,m,d,p,q,r,t)}$\\
    \hline

    \end{tabular}
    \label{tab:input_parameters_earth}
    
    {\raggedright \textbf{References.} (a)~\cite{Sotin2007}; (b)~\cite{Siebert2011}; (c)~\cite{Siebert2012}; (d)~\cite{Fischer2015}; (e)~\cite{Wade2005}; (f)~\cite{Li1996}; (g)~\cite{Wood2006}; (h)~\cite{Lee2009}; (i)~\cite{Stixrude2014}; (j)~\cite{Palme2014}; (k)~\cite{Lodders2019}; (l)~\cite{Zhang2016}; (m)~\cite{Wood2006}; (n)~\cite{Stewart2007}; (o)~\cite{Seagle2006}; (p)~\cite{Tsuno2013}; (q)~\cite{Ricolleau2011}; (r)~\cite{Rubie2011}; (s)~\cite{Lay2008}; (t)~\cite{Hirose2021}
    \\
    $\star$ Al, Ca, and Ni replaced by Si, Mg, and Fe (see \cite{Sotin2007})\par}
  \end{center}
\end{table}s

\begin{table}[ht]
  \begin{center}
    \caption{Permissible ranges for the output parameters for the Earth reference model.}
    \begin{tabular}{lccc}
      \hline
       Parameter & Unit & Model range & Literature  \\
      \hline
         $\rm Si\#$ &  & 0.47--0.52  & 0.47-0.56 $^{(d)\star}$ \\
         $X_{\rm FeO}^ {\rm Mantle}$ & mol\%  & 3.4-7.2 & 5.4-6 $^{(e,l)\star}$ \\
         $X_{\rm SiO_2}^ {\rm Mantle}$ & mol\% & 44-50 & 41-43 $^{(e,l)\star}$ \\
         $T_{\rm CS}$ & K & 3030-3690 & 2400-4200 $^{(a, b,h,i,j,k)}$\\
         $R_{\rm Core}$ & km & 3390-3620 & 3485$^{(n)}$\\
         $R_{\rm ICB}$ & km & 1040-1430  & 1215$^{(c, n)}$ \\
         $-\log(f_{\rm O_2})$ &   & 1.8--2.8  & 0.6-2.3 $^{(a, b, m)}$ \\
    \hline
    \end{tabular}
    \label{tab:output_parameters_earth}
    
    {\raggedright \textbf{References.} (a)~\cite{Siebert2011}; (b)~\cite{Siebert2012}; (c)~\cite{Monnereau2010}; (d)~\cite{Sotin2007}; (e)~\cite{Allegre1995}; (f)~\cite{Marty20120}; (g)~\cite{Ohtani2019}; (h)~\cite{Fischer2015}; (i)~\cite{Wade2005}; (j)~\cite{Li1996}; (k)~\cite{Wood2006}; (l)~\cite{Workman2005}; (m)~\cite{Doyle2019}; (n)~\cite{Jordan1974} \\
    $\star$ Al, Ca, and Ni replaced by Si, Mg, and Fe (see \cite{Sotin2007})\par}
  \end{center}
\end{table}

\end{document}